\documentclass[12pt]{article}

\usepackage{epsfig}
\usepackage{textcomp}
\newcommand{\Li}{\mbox{Li}_2}
\newcommand{\slp}{p \hspace{-1.6mm} /}
\newcommand{\sls}{s \hspace{-1.6mm} /}
\newcommand{\1}{1 \hspace{-0.95mm} {\rm l}}
\newcommand{\pp}{\phantom{-}}
\newcommand{\pf}{\protect \boldmath}
\newcommand{\LO}{{\sl LO }}
\newcommand{\NLO}{{\sl NLO }}
\newcommand{\NNLO}{{\sl NNLO }}
\newcommand{\pol}{\uparrow}
\newcommand{\decay}{\longrightarrow}
\newcommand{\GeV}{{\rm\,GeV}}

\newenvironment{Eqnarray*}{\arraycolsep 0.14em \begin{eqnarray*}}
{\end{eqnarray*}}
\renewcommand{\theequation}{\mbox{\arabic{equation}}}
\newcounter{saveeqn}
\newcommand{\alpheqn}{\setcounter{saveeqn}{\value{equation}}
\stepcounter{saveeqn}\setcounter{equation}{0}%
\renewcommand{\theequation}{\mbox{\arabic{saveeqn}\alph{equation}}}}
\newcommand{\reseteqn}{\setcounter{equation}{\value{saveeqn}}%
\renewcommand{\theequation}{\mbox{\arabic{equation}}}}

\begin{document} %%%%%%%%%%%%%%%%%%%%%%%%%%%%%%%%%%%%%%%%%%%%%%%%%%%%%%%%%%%%%%%

 \thispagestyle{empty}
 \begin{flushright}
     MZ-TH/03-09\\
     hep-ph/0306082\\
     June 2003\\
 \end{flushright}
 \vspace{5mm}

 \begin{center}
   {\Large \bf One--loop corrections to polarization observables}\\[7mm]
   {\large J.G.~K\"orner\footnotemark{} and M.C.~Mauser}\\[13mm] 
   {Institut f\"ur Physik, Johannes Gutenberg--Universit\"at}\\[2mm]
   {Staudinger Weg 7, D--55099 Mainz, Germany}
 \end{center}
 \vspace{15mm}
 
 \begin{abstract}
  We review the physics of polarization observables in
  high energy reactions in general and discuss the status of
  \NLO one--loop corrections to these observables in specific.
  Many high order radiative corrections exists for rates but not
  many \NLO radiative corrections exist for polarization observables.
  The radiative correction calculations for polarization observables
  are somewhat more complicated than those for rates.
  They tend to be smaller than those for the rates.
  In most of the examples we discuss we include mass effects which
  significantly complicate the radiative correction calculations.
  We elaborate a general scheme which allows one to enumerate the
  number of independent density matrix elements in a reaction and
  provide explicit examples of angular decay distributions in
  self--analyzing decays that allow one to experimentally measure the
  density matrix elements.
  We provide examples of reactions where certain density matrix elements
  are only populated at \NLO or by mass effects.
  In our discussion we concentrate on semileptonic bottom and
  top quark decays which are linked to leptonic $ \mu $ and $ \tau $
  decays through a Fierz transformation.
 \end{abstract}
 \vspace{10mm}

 \begin{center}
  {\sl $ {}^{1} $ \hspace{-2mm}
   Invited lecture at the International School ``Heavy Quark Physics'',\\
   Dubna, Russia, 27 May -- 5 June 2002\\
   To appear in the Proceedings (Lecture Notes in Physics)}
 \end{center}
 
 \pagebreak

%%%%%%%%%%%%%%%%%%%%%%%%%%%%%%%%%%%%%%%%%%%%%%%%%%%%%%%%%%%%%%%%%%%%%%%%%%%%%%%%
%
%                          Introduction 
%
%%%%%%%%%%%%%%%%%%%%%%%%%%%%%%%%%%%%%%%%%%%%%%%%%%%%%%%%%%%%%%%%%%%%%%%%%%%%%%%%

 \section{Introduction}

 Most of the \NLO radiative corrections to rates
 have been done quite some years ago.
 In fact, two--loop \NNLO calculations for rates are
 now becoming quite standard.
 In $ e^+ e^- $--annihilation one is even pushing hard to determine
 the {\sl NNNNLO} corrections to the $ R $ ratio \cite{bck02,bck02tau}.
 Contrary to this, many \NLO radiative correction calculations to
 polarization observables involving also massive quarks
 have only been done in the last few years.

 It is clear that higher order radiative corrections to
 unpolarized observables will always be at the center of attention
 because data on unpolarized observables are as a rule
 much more accurate than data for polarized observables.
 Nevertheless, as experimental data on polarization observables
 has been accumulating over the years there is an evolving need
 for radiative corrections to polarization observables.
 One of the reasons that polarized radiative correction were
 lagging behind unpolarized radiative corrections is that
 the computational effort in the calculation of radiative corrections
 to polarization observables is larger than that for rates.
 For once, one cannot sum over the spins of intermediate
 states whose polarization one wants to calculate.
 One therefore cannot make use of the powerful unitarity method to calculate
 radiative corrections from the absorptive parts of higher order loop graphs.
 Further, the definition of polarization observables brings in extra momentum
 factors in the integrands of the requisite phase space integrals which makes
 life more difficult.
 This is particularly true if the masses of particles
 in the process cannot be neglected.  
 For example, in the process $ t \rightarrow b + W^+ $ the
 longitudinal component of the polarization vector of the top
 quark along the $ W $--direction is given by  

 \begin{equation} %% def. of long. polariztion vector
 \label{polvektor1} 
   s_t^{l, \mu} = \frac{1}{|\vec{q}|} 
   \bigg( q^{\mu} - \frac{p_t \!\cdot\! q}{m_t^2} p_t^{\mu} \bigg),
 \end{equation}

 \noindent where the denominator factor
 $ |\vec{q}| = \sqrt{q_0^2 - m_W^2} $ comes in for
 normalization reasons ($ q $ is the momentum of the $ W^+ $).
 It is quite clear that such square root factors lead to
 nontrivial complications in the phase space integrations.
 Similar square root factors appear when projecting onto the
 polarization states of the $ W^+ $.

 Inclusion of mass effects as e.g. in the semileptonic decay
 $ b \rightarrow c + l^- + \bar{\nu}_l $
 ($ m_c / m_b \approx 0.30 $) or in the leptonic decay
 $ \tau^- \rightarrow \mu^- + \nu_{\tau} + \bar{\nu}_{\mu} $ 
 ($ m_{\mu} / m_{\tau} \approx 0.07 $) render the
 analytical calculations considerably more complicated.
 The villain is the K\'allen function
 $ (m_1^4 + m_2^4 + m_3^4 - 2 (m_1^2 m_2^2 + m_1^2 m_3^2 + m_2^2 m_3^2))^{1/2} $
 which is brought in by extra three--momentum factors.
 In the case that e.g. $ m_3 \rightarrow 0 $
 the K\'allen function simplifies to $ (m_1 - m_2)^2 $ which leads
 to an enormous simplification in the phase space integrals.

\begin{figure}[htbp] %% plot I
   \begin{center} %% long. pol. of the electron
     \includegraphics[width=.60 \textwidth, clip=]{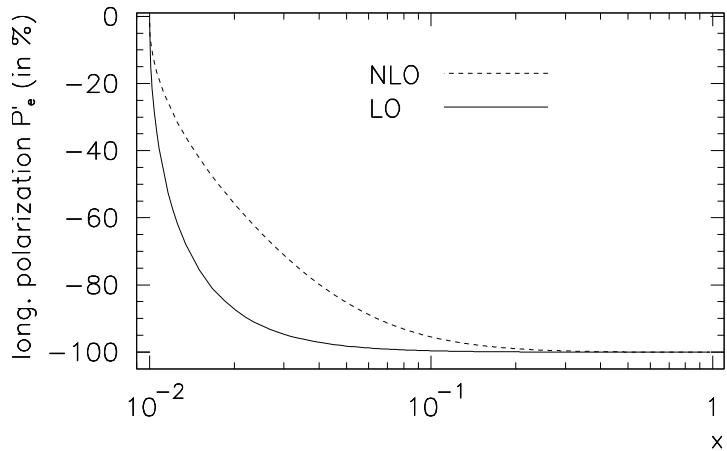}
   \end{center}
   \caption[]{Longitudinal polarization of the electron
    in leptonic muon decays at \LO and \NLO as a function of the
    scaled energy $ x = 2 E_e / m_{\mu} $ \cite{fgkm03electron}.
    The \LO curve is very well described by the
    functional behaviour $ P^l_e = -\beta $.}
   \label{plot1}
 \end{figure}

 Physically speaking mass effects become large in regions of
 phase space where the massive particles become nonrelativistic.
 For example, in the leptonic decay of the muon  
 $ \mu^- \rightarrow e^- + \nu_{\mu} + \bar{\nu}_e $,
 the mass of the electron cannot be neglected in the
 threshold region where the energy of the electron is small.
 This is illustrated in Fig.~\ref{plot1} where the longitudinal
 polarization of the electron is plotted against the scaled energy
 $ x = 2 E_e / m_{\mu} $ of the electron \cite{fgkm03electron}.
 We have chosen a logarithmic scale for $ x $
 in order to enhance the threshold region. 
 In the threshold region the longitudinal polarization
 deviates considerably from the naive value $ P^l_e = -1 $.
 The radiative corrections in the threshold region
 can be seen to be quite large. 

 All our results are given in closed analytic form.
 In the days of fast numerical computers one may rightfully ask
 what the advantage of having closed form expressions is.
 One of the advantages is that all requisite mass and
 momentum limits can be taken in analytic form.
 Thus the analytic results can be subjected to tests against known
 limiting cases providing for the necessary checks of the full results.
 A second issue is user--friendliness.
 The analytic expressions are simple enough to be incorporated
 in numerical programs by the prospective user.
 Parameter values such as masses and coupling constants can be varied
 at will by the user in his own program without having to refer to 
 numerical programs written by others.
 We have checked that our analytical formulas are numerically stable
 even in the small and large mass limits.

 \NLO corrections to rates can be quite large.
 \NLO corrections to polarization observables have
 a tendency to be somewhat smaller.
 The reason is that polarization observables are normalized quantities.
 They are normalized with regard to the total rate.
 The numerators and denominators in the relevant polarization expressions
 tend to go in the same direction.
 The reason is that the \NLO numerators and denominators are dominated by the
 soft gluon or soft photon contributions which are universal in the sense
 that they multiply the relevant Born term expressions and thus cancel out
 in the ratio.
 For example, for $ t \rightarrow b + W^+ $ the $ O(\alpha_s) $ correction
 to the rate is $ - 8.5 \% $ and $ (- 2.5 \% \div + 3 \%) $ for the
 polarization observables \cite{fgklm99,fgkm01,fgkm02}.

 At \NLO some density matrix elements become populated which vanish at
 leading order.
 An example is again the decay $ t \rightarrow b + W^+ $ where the
 rate into the right--handed $ W^+ $ vanishes ($ \Gamma(W_R) = 0 $) for 
 $ m_b = 0 $ at \LO but where $ \Gamma(W_R) \ne 0 $ at {\sl NLO}.
 Generally speaking, Standard Model radiative corrections can populate
 density matrix elements elements that vanish at leading order.
 Consequently radiative corrections change angular decay distributions
 as do new physics effects.
 The lesson to be learned is clear.
 Before ascribing a given polarization effect to new physics one has to make
 sure that it does not result from radiative corrections of old physics.
 In the above example a non--zero $ \Gamma(W_R) \ne 0 $ could result from
 an admixture of a right--handed charged current (new physics) or from
 radiative corrections.
 Only the precise knowledge of the magnitude of the radiative corrections
 allows one to exclude or conclude for new physics effects.

 In this review we will mostly be concerned with radiative corrections
 to the current--induced transitions $ t \rightarrow b $, $ b \rightarrow c $
 and $ l \rightarrow l' $.
 It should be clear that the additional gluons (or photons) in
 radiative correction calculations only couple to the
 $ q_1 \rightarrow q_2 $ (or $ l \rightarrow l' $) side of the
 relevant semileptonic (or leptonic) transitions.
 This implies that the structure of the \NLO radiative corrections
 is the same in the two classes of processes.
 The current--induced transition $ l \rightarrow l' $ is not in the
 Standard Model form but in the charge retention form.
 However, the charge retention form can be linked to the
 Standard Model form by the remarkable property that the
 $ (V - A)^\mu (V - A)_\mu $ interaction is an eigenvector
 under Fierz crossing (see Sect.~3).
 One therefore has

 \begin{eqnarray} %% charge retention
   \label{chargeretention}
   {\cal L} & = & \frac{G_F}{\sqrt{2}} 
   [\bar{\nu}_{\mu} \gamma^{\alpha} (\1 - \gamma_5) \mu] 
   [\bar{e} \gamma_{\alpha} (\1 - \gamma_5) \nu_{e}] + \mbox{h.c.} \\[2mm]
            & = & \frac{G_F}{\sqrt{2}} 
   [\bar{\nu}_{\mu} \gamma^{\alpha} (\1 - \gamma_5) \nu_{e}] 
   [\bar{e} \gamma_{\alpha} (\1 - \gamma_5) \mu ] + \mbox{h.c.}
 \end{eqnarray}

 \noindent All of the above three transitions are therefore governed
 by the same matrix elements.
 In particular the \NLO QED and QCD radiative corrections have the
 same structure for all three transitions.
 In Table~1 we provide a list of the processes for which the
 radiative corrections to the rates and the polarization observables
 can all be calculated from essentially the same set of \NLO matrix elements.
 In this review we shall only present a few sample results from the
 processes in Table~\ref{Tab1} because the results are to numerous to
 fit into a review of the present size.
 Instead we attempt to share our insights into the general features of
 spin physics and illustrate these with sample results taken from
 the processes in Table~\ref{Tab1}.  

 \begin{table} %% Table I
 \caption{List of processes based on the same current--induced matrix elements}
   \begin{center}
   \renewcommand{\arraystretch}{1.4}
   \setlength\tabcolsep{5pt}
   \begin{tabular}{lll}
     \hline\noalign{\smallskip}
       parton level \hspace{20mm} &
       particle level \hspace{10mm} &
       references \hspace{10mm}\\
     \noalign{\smallskip}
     \hline
     \noalign{\smallskip}
      $ t(\pol) \decay b + W^+(\pol) $ &
      $ t(\pol) \decay X_b + W^+(\pol) (\decay l^+ + \nu_l) $ &
      \cite{fgklm99,fgkm01,fgkm02,dgkm03}\\
      $ t(\pol) \decay b + H^+ $ &
      $ t(\pol) \decay X_b + H^+ $ &
      \cite{km03}\\
      $ b(\pol) \decay c(\pol) + l^- + \bar{\nu}_l $ &
      $ B \decay X_c + D_s^{(*)}(\pol) $ &
      \cite{fgkm00}\\
      $ $ &
      $ \Lambda_b(\pol) \decay X_c + D_s^{(*)}(\pol) $ & 
      \cite{fgkm03lambda}\\
      $ b(\pol) \decay u(\pol) + l^- + \bar{\nu}_l $ &
      $ B \decay X_u + \pi(\rho (\uparrow)) $ &
      \cite{fgkm00}\\
      $ $ &
      $ \Lambda_b(\pol) \decay X_u + \pi(\rho (\pol)) $ & 
      \cite{fgkm03lambda}\\  
      $ l^-(\pol) \decay  l'^-(\pol) + \nu_l + \bar{\nu}_{l'} $ &  & 
      \cite{fgkm03electron}\\ 
      $ (l,l')=(\mu,e),(\tau,\mu),(\tau,e) $ &\\[2mm]
      \hline
   \end{tabular}
   \end{center}
 \label{Tab1}
 \end{table}

%%%%%%%%%%%%%%%%%%%%%%%%%%%%%%%%%%%%%%%%%%%%%%%%%%%%%%%%%%%%%%%%%%%%%%%%%%%%%%%%
%
%          Miscellaneous Remarks on Polarization Effects 
%
%%%%%%%%%%%%%%%%%%%%%%%%%%%%%%%%%%%%%%%%%%%%%%%%%%%%%%%%%%%%%%%%%%%%%%%%%%%%%%%%

\section{Miscellaneous Remarks on Polarization Effects}

%%%%%%%%%%%%%%%%%%%%%%%%%%%%%%%%%%%%%%%%%%%%%%%%%%%%%%%%%%%%%%%%%%%%%%%%%%%%%%%%
%
%                      Examples of 100 \% polarization 
%
%%%%%%%%%%%%%%%%%%%%%%%%%%%%%%%%%%%%%%%%%%%%%%%%%%%%%%%%%%%%%%%%%%%%%%%%%%%%%%%%

\subsection{Examples of \pf $ 100 \% $ polarization}

 In this subsection we discuss examples of $ 100 \% $ polarization.
 Cases of $ 100 \% $ polarization are usually associated with limits
 when masses or energies of particles become small or large compared
 to a given scale in a process.
 For example, an on--shell or off--shell gauge bosons radiated off massless
 fermions (leptons or quarks) is purely transverse since fermion helicity is
 conserved by the vector and axial vector Standard Model couplings of
 the gauge bosons.
 Contrary to this, massive gauge bosons become purely longitudinal in
 the high energy limit (see Sect.~11).
 We discuss two examples of $ 100 \% $ polarization in more detail.
 Namely, the case of a left--chiral fermion which becomes purely left--handed 
 in the chiral limit when $ m_f \rightarrow 0 $ and the case of soft gluon
 radiation from heavy quarks.

 \subsection*{Relativistic left--chiral fermions}

 Consider a fermion moving along the $ z $--axis.
 The positive and negative helicity polarization four--vectors of the fermion
 are given by $ s_{\pm}^{\mu} = \pm (p; 0, 0, E)/m $.
 At first sight it appears to be problematic to take the limit
 $ m \rightarrow 0 $ in expressions involving the polarization
 four--vectors because of the denominator mass factor.
 However, the saving feature is that $ s_{\pm}^{\mu} $ becomes increasingly
 parallel to the momentum $ \pm p^\mu $ when $ m \rightarrow 0 $.
 In fact, one has 

 \begin{equation}
   s_{\pm}^{\mu} = \pm p^{\mu} + O(m/E),
 \end{equation}

 \noindent since $ E = p + (m^2)/(E + p) $.
 Therefore the projectors onto the two helicity states simplify to 

 \begin{equation}
   \label{casimir}
   u(\pm) \bar{u}(\pm) =
   \frac{1}{2} (\slp + m) (\1 + \gamma_5 \sls_{\pm})
   \stackrel{m \rightarrow 0} {- \!\!\! - \!\!\! \longrightarrow}
   \frac{1}{2} \slp(\1 \mp \gamma_5).
 \end{equation}
 
 This shows that a right/left--chiral fermion is purely
 right/left--handed in the chiral limit.
 Thus the final state electron emerging from leptonic $ \mu $--decay
 is purely left--handed in the chiral limit except for the anomalous
 spin--flip contribution appearing first at \NLO which populate also
 the right--handed state.
 The anomalous spin--flip contribution will be discussed in Sect.~12.

 The approach to the chiral limit is well described by an
 approximate formula frequently discussed in text books.
 The argument goes as follows.

 Introduce left--chiral fermion spinors according to 
 $ u_L(\pm) = 1/2 (\1 - \gamma_5) u(\pm) $,
 where $ u(\pm) $ are helicity $ \lambda = +1/2, -1/2 $ spinors.
 The longitudinal polarization of a left--chiral electron is then
 calculated by taking the ratio of the difference and the sum
 of the $ \lambda = +1/2 $ and $ \lambda = -1/2 $ scalar
 densities according to
 
 \begin{eqnarray}
   \label{minusbeta}
   P^l_e & = & \frac{
   u^{\dagger}_{L}(+) u_{L}(+) -
   u^{\dagger}_{L}(-) u_{L}(-)}{
   u^{\dagger}_{L}(+) u_{L}(+) +
   u^{\dagger}_{L}(-) u_{L}(-)} \\[2mm] & = &
   \frac{(E + m -p)^2 - (E + m + p)^2}
   {(E + m - p)^2 + (E + m + p)^2} = - \frac{p}{E} = - \beta. 
 \end{eqnarray}

 \noindent A corresponding result holds for
 left--chiral positrons where $ P^l_e = \beta $.

 The accuracy of the above result can be checked with
 the corresponding expression for the Born term
 polarization of the electron in leptonic $ \mu $--decay.
 One finds (see e.g. \cite{fgkm03electron})

 \begin{equation}
   P^l_e = - \beta \frac{x (3 - 2 x + y^2)}
   {x (3 - 2 x) - (4 - 3 x) y^2},
 \end{equation}

 \noindent where $ x = 2 E_e / m_{\mu} $ and $ y = m_e / m_{\mu} $.
 For the Born term contribution the correction to the approximate result
 (\ref{minusbeta}) $ P^l_e = - \beta $ is of $ O(1 \%) $ or less such that
 the correct and approximate curves are not discernible at the scale of
 Fig.~\ref{plot1}.
 Contrary to this the \NLO polarization deviates substantially from a simple
 $ P^l_e = -\beta $ behaviour (see Fig.~\ref{plot1}) even though one has
 again $ P^l_e \propto -\beta $ at \NLO \cite{fgkm03electron}.
 This may have to do with the fact that the radiative corrections
 involve an extra photon emission from the electron leg.
 The electron is therefore no longer left--chiral at \NLO
 as in the \LO Born term case. 

%%%%%%%%%%%%%%%%%%%%%%%%%%%%%%%%%%%%%%%%%%%%%%%%%%%%%%%%%%%%%%%%%%%%%%%%%%%%%%%%
%
%          Soft gluon or soft photon radiation 
%
%%%%%%%%%%%%%%%%%%%%%%%%%%%%%%%%%%%%%%%%%%%%%%%%%%%%%%%%%%%%%%%%%%%%%%%%%%%%%%%%

 \subsection*{Soft gluon or soft photon radiation}

 When a soft gluon (or a soft photon) is radiated off a fermion line
 it is $ 100 \% $ polarized in the plane spanned by the fermion and
 the gluon (or photon).
 To see this in an exemplary way consider the tree graph matrix element
 of a soft gluon radiated off a top or antitop in the process
 $ e^+ e^- \rightarrow t \bar{t} g $.
 The soft gluon matrix element reads

 \begin{equation}
   \label{softgluon} %% matrix element
   T_{\mu}^{\mathrm{(s.g.)}}(\pm) =
   T_{\mu}^{\mathrm{(Born)}} 
   \left( \frac{p_t^{\alpha}}{p_t \!\cdot\! k} -
   \frac{p_{\bar{t}}^{\alpha}}{p_{\bar{t}} \!\cdot\! k} \right)
   \epsilon^{\ast}_{\alpha} (\pm)
 \end{equation}

 \noindent with $ \epsilon^{\ast}_{\alpha} (\pm) = (0; \mp 1, i, 0)/\sqrt{2} $.

 The $ 2 \times 2 $ density matrix of the gluon can be expanded
 in terms of the unit matrix $ \1 $ and the Pauli matrices
 $ \sigma_i $ according to

 \begin{equation} %% density matrix
   \left( \begin{array}{cc}
   h_{+} h_{+}^{*} \: & h_{+} h_{-}^{*} \\[1mm]
   h_{-} h_{+}^{*} \: & h_{-} h_{-}^{*} 
   \end{array} \right) =
   \frac{1}{2} (\sigma \!\cdot\! \1 +
   \vec{\xi} \!\cdot\! \vec{\sigma}),
 \end{equation}

 \noindent where $ \vec{\xi} $ denotes the Stokes ``vector''.
 We have set ``vector'' in quotation marks because the Stokes ``vector''
 does not transform as a vector but rather as a spin $ 2 $ object under
 three--dimensional rotations.
 From (\ref{softgluon}) one sees that the helicity amplitudes are
 relatively real and that $ h_{+} = - h_{-} $.
 The normalized Stokes ``vector'' is then given by  

 \begin{equation} %% normalized Stokes vector
   \vec{\xi} / \sigma = (-1, 0, 0),
 \end{equation}

 \noindent i.e., in the terminology of classifying the polarization states
 of a gluon (or a photon), the polarization of the gluon is $ 100 \% $
 linearly polarized in the production plane.

 \begin{figure}[hbtp] %% plot II
   \begin{center} %% linear polarization
     \includegraphics[width=.60 \textwidth, clip=]{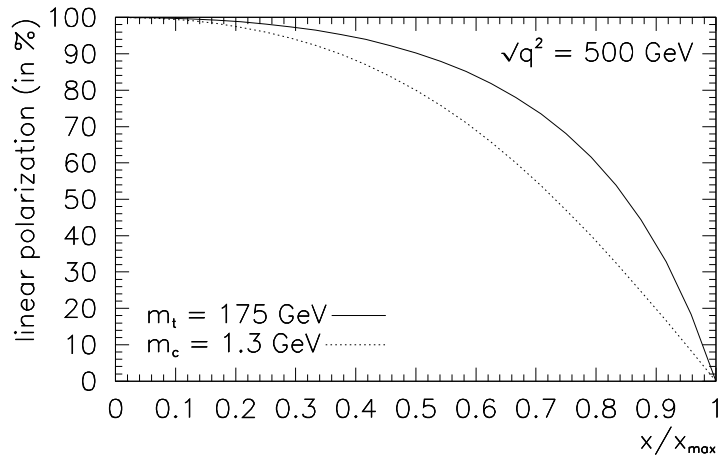}
   \end{center}
   \caption[]{Linear polarization of the gluon as a
    function of the scaled energy of the gluon in
    $ e^+ e^- \rightarrow  Q \, \bar{Q} \,g $ \cite{gkl97}.}
   \label{plot2}
 \end{figure}

 In Fig.~\ref{plot2} we show a plot of the linear polarization of
 the gluon as a function of the scaled energy of the gluon for
 $ e^+ e^- \rightarrow t \, \bar{t} \, g $ and for
 $ e^+ e^- \rightarrow c \, \bar{c} \, g $ \cite{gkl97}.
 At the soft end of the spectrum the gluon is 
 $ 100 \% $ polarized and then slowly drops to zero polarization
 at the hard end of the spectrum.
 It is quite remarkable that a high degree of polarization is
 maintained over a large part of the spectrum.   

%%%%%%%%%%%%%%%%%%%%%%%%%%%%%%%%%%%%%%%%%%%%%%%%%%%%%%%%%%%%%%%%%%%%%%%%%%%%%%%%
%
%                   Examples of zero polarization 
%
%%%%%%%%%%%%%%%%%%%%%%%%%%%%%%%%%%%%%%%%%%%%%%%%%%%%%%%%%%%%%%%%%%%%%%%%%%%%%%%%

 \subsection{Examples of zero polarization}

 The mean of a single spin polarization observable such as
 $ \langle \vec{\sigma} \!\cdot\! \vec{p} \rangle $ is a parity--odd measure.
 In strong interaction processes, which are parity conserving,
 there are no single spin polarization effects, i.e.
 $ \langle \vec{\sigma} \cdot \vec{p} \rangle = 0 $
 (except for the $ T $--odd single spin effects
 $ \propto \langle \vec{\sigma} \cdot (\vec{p_i} \times \vec{p_j}) \rangle $
 to be discussed later on).
 This is different in weak and electroweak interactions where the presence of
 parity violations induces a wide variety of single spin polarization phenomena.
 As an example we take $ e^+ e^- $--annihilation into heavy quarks where single
 spin polarization phenomena occur due to $ (\gamma,Z) $--exchange.
 \NLO corrections to these single spin polarization phenomena
 have been discussed in \cite{Korner:1993dy,Groote:1995yc,Groote:1995ky,Groote:1996nc,Brandenburg:1998xw,Ravindran:2000rz}.
 Another example is the semileptonic decay of a polarized heavy quark where
 the \NLO corrections to the lepton spectrum have been presented in
 \cite{Czarnecki:1990pe,Czarnecki:gt}.

 However, there are double spin polarization effects in strong interactions
 since $ \langle \vec{\sigma} \!\cdot\! \vec{\sigma} \rangle $ is a parity even
 measure.
 An example is the correlation between the spins of the top
 and the antitop produced in hadronic collisions
 \cite{Brandenburg:1996df,Mahlon:1997uc,Bernreuther:2001rq,Bernreuther:2001bx}.
 Naturally, double spin polarization effects also occur in weak interaction
 processes such as in high energy $ e^+ e^- $--annihilation 
 (see \cite{Brandenburg:1998xw,Parke:1996pr,Groote:1997su,Tung:1997ur}). 
 However, even in weak processes single spin polarization effects can
 vanish when one approaches phase space boundaries.
 Consider, for example, the process $ e^+ e^- \rightarrow t \, \bar{t} $
 at threshold.
 In general, the vector current induced amplitude is $ s $-- and $ d $--wave,
 while the axial vector current amplitude is $ p $--wave.
 As one approaches threshold the $ s $--wave amplitude will dominate and
 thus there will be no vector--axial vector interference.
 Therefore single top polarization goes to zero
 in $ e^+ e^- \rightarrow t \, \bar{t} $ as one
 approaches the threshold region.
 
 Another example of zero polarization are \NLO $ T $--odd measures
 in the process  $ e^+ e^- \rightarrow q \, \bar{q} \, (g) $
 for massless quarks.
 $ T $--odd effects arise from the imaginary parts of one--loop contributions.
 However, in the case $ e^+ e^- \rightarrow q \, \bar{q} \, (g) $ the imaginary
 parts can be shown to be proportional to the Born term contribution in the
 mass zero limit.
 Hence, the $ T $--odd measures are zero in this case \cite{Korner:1980np}.    
 When $ m_q \ne 0 $ the imaginary parts are no longer proportional
 to the Born term and, consequently, one obtains nonzero values for the
 $ T $--odd measures \cite{Fabricius:1980wg,Brandenburg:1995nv}.
 In the crossed channels (deep inelastic scattering, Drell--Yan process)
 the imaginary parts no longer have Born term structure and one obtains
 non--vanishing $ T $--odd effects even for zero mass quarks
 \cite{Hagiwara:1981qn,Hagiwara:1982cq,Hagiwara:1984hi,Korner:2000zr}.

%%%%%%%%%%%%%%%%%%%%%%%%%%%%%%%%%%%%%%%%%%%%%%%%%%%%%%%%%%%%%%%%%%%%%%%%%%%%%%%%
%
%                                Mass effects 
%
%%%%%%%%%%%%%%%%%%%%%%%%%%%%%%%%%%%%%%%%%%%%%%%%%%%%%%%%%%%%%%%%%%%%%%%%%%%%%%%%

 \subsection{Mass effects}

 Mass effects make the radiative correction more complicated.
 We attempt to illustrate the complication brought about by an additional
 mass scale by listing the ({\sl LO} + {\sl NLO}) results for the total rate
 of leptonic $ \mu $ decays $ \mu^- \rightarrow e^- + \bar{\nu}_e + \nu_{\mu} $
 for (i) $ y = m_e / m_{\mu} = 0 $ and for (ii) $ y = m_e / m_{\mu} \ne 0 $.
 In case (i) one has

 \begin{equation} %% rate massless
   \label{ratemassless}
   \Gamma = \frac{G_F^2 m_{\mu}^5}{192 \pi^3}
   \left( 1 + \frac{\alpha}{\pi}
   \bigg\{ \frac{25 - 4 \pi^2}{8} \bigg\} \right).
 \end{equation}

 \noindent For case (ii) one obtains

 \begin{eqnarray} %% massive rate
   \label{ratemass}
   \Gamma & = & \frac{G_F^2 m_{\mu}^5}{192 \pi^3}
   \bigg( (1 - y^4)(1 - 8 y^2 + y^4) - 12 y^4 \ln(y) +
   \frac{\alpha}{\pi} \bigg\{ \nonumber \\ & + &
   \frac{1}{24} (1 - y^4) (75 - 956 y^2 + 75 y^4) -
   2 y^4 (36 + y^4) \ln^2(y) \nonumber \\ & - &
   \frac{\pi^2}{2} (1 - 32 y^3 +
   16 y^4 - 32 y^5 + y^8) \nonumber \\ & - &
   \frac{1}{3} (60 + 270 y^2 - 4 y^4 + 17 y^6) y^2
   \ln(y) \nonumber \\ & - &
   \frac{1}{6} (1 - y^4)(17 - 64 y^2 + 17 y^4)
   \ln(1 - y^2) \nonumber \\[2mm] & + &
    4 (1 - y)^4 (1 + 4 y + 10 y^2 + 4 y^3 + y^4)
   \ln(1 - y) \ln(y) \nonumber \\[3mm] & + &
    4 (1 + y)^4 (1 - 4 y + 10 y^2 - 4 y^3 + y^4)
   \ln(1 + y) \ln(y) \nonumber \\[3mm] & + &
   2 (3 + 32 y^3 + 48 y^4 + 32 y^5 + 3 y^8)
   \Li(-y) \nonumber \\[3mm] & + &
   2 (3 - 32 y^3 + 48 y^4 - 32 y^5 + 3 y^8)
   \Li(y) \bigg\} \bigg).
 \end{eqnarray}

 The length of the \NLO term in (\ref{ratemass}) illustrates but does not
 describe the added  complication when introducing a new mass scale.
 The inclusion of full mass effects is not of much relevance for leptonic
 $ \mu^- \rightarrow e^- $ decays (except in the threshold region)
 where $ (m_e / m_{\mu})^2 = 2.34 \cdot 10^{-5} $ but may be relevant
 for the leptonic $ \tau \rightarrow \mu $ decays, where e.g.
 $ (m_{\mu} / m_{\tau})^2 = 3.54 \cdot 10^{-3} $.
 In fact, experimental data on leptonic $ \tau $ decays do show a
 mass dependence of the partial rates on the daughter lepton's mass.
 For the decays $ \tau^- \rightarrow \mu^- + \bar{\nu}_{\mu} + \nu_{\tau} $ and 
 $ \tau^- \rightarrow e^- + \bar{\nu}_e + \nu_{\tau} $ one finds branching
 ratios of $ 17.37 \pm 0.06 \% $ and  $ 17.84 \pm 0.06 \% $ \cite{pdg02}.
 In agreement with experiment the \LO result predicts a reduction of the
 partial rate by $ 2.82 \% $ for the $ \tau^- \rightarrow \mu^- $ mode
 relative to the $ \tau^- \rightarrow e^- $ mode.
 As concerns \NLO effects, the experimental errors on the two partial rates
 are still too large to allow for checks on the mass dependence of the
 \NLO result.
 Quite naturally, for semileptonic $ b \rightarrow c $ decays,
 where $ y = m_c / m_b \approx 0.3 $, the mass dependence of the daughter
 quark must be kept.

\begin{figure}[htbp] %% plot III
   \begin{center} %% decay L/R
     \includegraphics[width=.65 \textwidth]{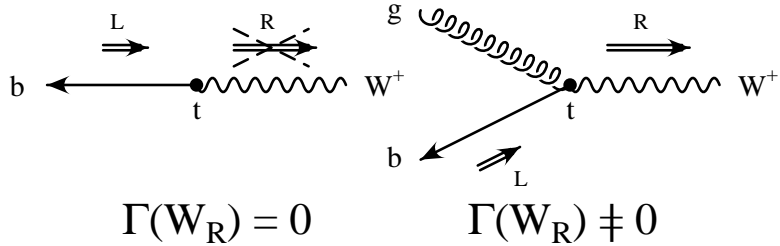}
   \end{center}
   \caption[]{The decay $ t \rightarrow b + W^+_R $ at \LO and \NLO}
   \label{plot3}
 \end{figure}

 Mass effects also populate density matrix elements which are zero
 when a given mass is set to zero.
 A prominent example is again $ e^+ e^- \rightarrow q \, \bar{q} $ which is
 purely transverse for mass zero quarks but acquires a longitudinal component
 when the quark becomes massive.
 In the same vein one has the Callan--Gross relation $ F_2 = x F_1 $
 in deep inelastic scattering for mass zero quarks which is spoiled
 when the quarks become massive.
 Another example is the density matrix of the
 electron in leptonic muon decay.
 Its off--diagonal elements contribute to the transverse component of the
 electron's  polarization (see (\ref{fermiondensity})).
 The off--diagonal elements are proportional to the positive helicity
 amplitude $ h_{1/2} $ which in turn is proportional to the mass of the
 electron  ($ h_{1/2} \sim m_e $).
 One therefore has $ P^{\perp}_e \sim m_e $ \cite{fgkm03electron}.
 A third example is the transition of the top quark into a right--handed
 $ W^+ $ ($ W_R $) and a bottom quark.
 For $ m_b = 0 $ the left--chiral bottom quark becomes purely
 left--handed and, from angular momentum conservation, one therefore finds
 $ \Gamma(W_R) = 0 $ (see Fig.~\ref{plot3}) (see e.g. \cite{fgkm01}).

%%%%%%%%%%%%%%%%%%%%%%%%%%%%%%%%%%%%%%%%%%%%%%%%%%%%%%%%%%%%%%%%%%%%%%%%%%%%%%%%
%
%        Some polarization measures are NLO effects 
%
%%%%%%%%%%%%%%%%%%%%%%%%%%%%%%%%%%%%%%%%%%%%%%%%%%%%%%%%%%%%%%%%%%%%%%%%%%%%%%%%

 \subsection{Some polarization measures are {\boldmath $ N\!LO $} effects}
 
 Whenever a given polarization component is zero at the \LO Born term
 level, and the vanishing of this polarization component is not due to
 general symmetry principles, it is very likely that this polarization
 component will be populated by gluon or photon emission at {\sl NLO}.
 We shall list a few examples where this is the case.
 
 A prominent example is the Callan--Gross relation
 which is violated by gluon emission at {\sl NLO}.
 Similarly, a longitudinal contribution is generated in
 $ e^+ e^- $--annihilation into massless quarks by gluon emission.
 T--odd measures derive from imaginary parts of one--loop contributions,
 i.e. they start only at {\sl NLO}.
 As mentioned before, a transverse--plus polarization of the
 $ W^+ $ in the decay $ t \rightarrow b + W^+ $ is generated by
 gluon emission as illustrated in Fig.~\ref{plot3}.
 At \LO the longitudinal polarization of massive quarks in 
 $ e^+ e^- \rightarrow  Q \bar{Q} $ is purely transverse, i.e.
 the polar distribution of this polarization component
 is proportional to $ (1 + \cos^2 \theta) $.
 The appearance of a longitudinal dependence proportional to
 $ \sin^2 \theta $ is a \NLO effect \cite{Groote:1995yc}.
 Anomalous helicity flip contributions involving massless
 quarks or leptons occur only at \NLO as discussed in Sect.~12. 

 We have already mentioned the $ T $--odd measures which,
 in a $ CP $ invariant theory, obtain contributions only from
 the imaginary parts of \NLO one--loop contributions.
 Apart from the examples listed in Sect.~2.2 we mention the case of
 the transverse polarization of a top quark produced in hadronic collisions
 \cite{Dharmaratna:jr,Bernreuther:1995cx,Dharmaratna:xd}.

%%%%%%%%%%%%%%%%%%%%%%%%%%%%%%%%%%%%%%%%%%%%%%%%%%%%%%%%%%%%%%%%%%%%%%%%%%%%%%%%
%
%        Presentation of NLO results of polarization observables 
%
%%%%%%%%%%%%%%%%%%%%%%%%%%%%%%%%%%%%%%%%%%%%%%%%%%%%%%%%%%%%%%%%%%%%%%%%%%%%%%%%

 \subsection{Presentation of {\boldmath $ N\!LO $}
  results of polarization observables}
 
 We have always chosen to present our analytical and numerical \NLO results
 on polarization observables in the form

 \begin{equation} 
   \label{polobs1}
   \langle P \rangle =
   \frac{N}{D} =
   \frac{N^{\mathrm{LO}}
   \big(1 + \alpha_s \hat{N}^{\mathrm{NLO}} \big)}
   {D^{\mathrm{LO}}
   \big(1 + \alpha_s \hat{D}^{\mathrm{NLO}} \big)},
 \end{equation}

 \noindent and not as 

 \begin{equation}
   \label{polobs2} 
   \langle P \rangle =
   \frac{N^{\mathrm{LO}}}
   {D^{\mathrm{LO}}}
   \big( 1 + \alpha_s (
   \hat{N}^{\mathrm{NLO}} -
   \hat{D}^{\mathrm{NLO}} ) \big).
 \end{equation}

 An argument for our preference of the form (\ref{polobs1}) over
 (\ref{polobs2}) is the following.
 It is common usage to try and extend a perturbation series beyond
 a given known order by Pad\'e improving the perturbation series.
 For example, for a second order perturbation series the
 Pad\'e $ P(1,1) $ improvement reads 

 \begin{equation} 
   \label{pade}
   1 + \alpha_s a_1 +
   \alpha_s^2 a_2 \approx
   \frac{1 + \alpha_s
   \frac{a_1^2 - a_2}{a_1}}
   {1 - \alpha_s \frac{a_2}{a_1}}.
 \end{equation}
 
 \noindent What has been done in the Pad\'e $ P(1,1) $ improvement is to fix
 the first order coefficients in the numerator and denominator of the r.h.s.
 of (\ref{pade}) by expanding the r.h.s. in powers of $ \alpha_s $ and then
 equating coefficients in (\ref{pade}).
 Note that at the \NLO level Pad\'e's method does not yet allow one to
 reconstruct (\ref{polobs1}) from (\ref{polobs2}).
 Considering the successes of the Pad\'e improvement program in
 other applications we nevertheless believe that the form
 (\ref{polobs1}) is a better approximation to the whole
 perturbation series than (\ref{polobs2}). 
 
 As an example where the use of the form (\ref{polobs2}) can lead to
 a misunderstanding is the \NLO result of the mean forward--backward asymmetry
 $ \langle A_{\mathrm{FB}} \rangle $ in $ e^+ e^- $--annihilation
 into a pair of massless quarks which, when using the form (\ref{polobs2}),
 is sometimes stated as

 \begin{equation} 
   \label{fb}
   \langle A_{\mathrm{FB}} \rangle =
   \frac{\sigma_{\mathrm{FB}}^{\mathrm{LO}}}
   {\sigma^{\mathrm{LO}}} (1 - \alpha_s / \pi ). 
 \end{equation}
 
 This result is very suggestive of a de facto non--vanishing
 radiative correction to $ \sigma_{\mathrm{FB}} $.
 However, the true result is $ \sigma_{\mathrm{FB}}^{\mathrm{NLO}} = 0 $
 \cite{Korner:1985dt}.
 The non--vanishing radiative correction to $ \langle A_{\mathrm{FB}} \rangle $
 in (\ref{fb}) is just a reflection of the well--known \NLO result for the rate 
 $ \sigma = \sigma^{\mathrm{LO}} (1 + \alpha_s / \pi) $ in the denominator of
 (\ref{polobs1}).
 To our knowledge the vanishing of $ \sigma_{\mathrm{FB}}^{\mathrm{NLO}} $
 evades a simple explanation and must be termed to be a dynamical accident.

%%%%%%%%%%%%%%%%%%%%%%%%%%%%%%%%%%%%%%%%%%%%%%%%%%%%%%%%%%%%%%%%%%%%%%%%%%%%%%%
%
%                Fierz Transformation 
%
%%%%%%%%%%%%%%%%%%%%%%%%%%%%%%%%%%%%%%%%%%%%%%%%%%%%%%%%%%%%%%%%%%%%%%%%%%%%%%%

 \section{Fierz Transformation}

 Fierz crossing exchanges two Dirac indices in the contracted product
 of two strings of $ \Gamma $--matrices according to ($ i $ not summed)
 
 \begin{equation}
   \big[ \Gamma_i^{\{ \mu \}} \big]_{\alpha \beta} 
   \big[ \Gamma_{i \{ \mu \}} \big]_{\gamma \delta} \rightarrow
   \big[ \Gamma_i^{\{ \mu \}} \big]_{\alpha \delta} 
   \big[ \Gamma_{i \{ \mu \}} \big]_{\gamma \beta},
 \end{equation}

 \noindent where the five currents in the set are conventionally
 labelled by $ i = S $, $ V $, $ T $, $ A $, $ P $ with
 $ \Gamma_i^{\{ \mu \}} = \big\{ 1 $, $ \gamma^\mu $,
 $ \sigma^{\mu \nu} $, $ \gamma^\mu \gamma_5 $, $ \gamma_5 \big\} $.
 The set of crossed configurations can then be expressed in terms of
 the set of uncrossed configurations.
 The five--by--five matrix relating the two sets is called the
 Fierz crossing matrix $ C_{Fierz} $.
 It is clear that one gets back to the original configuration
 when crossing twice, i.e. $ C_{Fierz}^2 = 1 $.
 The eigenvalues $ \lambda $ of the Fierz crossing matrix are thus
 $ \lambda = \pm 1 $ as can easily be seen by going to the
 diagonal representation.
 In explicit form the Fierz crossing matrix is given by 

 \begin{equation}
   \label{fierz}
   C_{Fierz} = \frac{1}{4} \left(
   \begin{array}{r@{\quad}r@{\quad}c@{\quad}r@{\quad}r}
     1 &  1 & 1/2 & -1 &  1 \\[1mm]
     4 & -2 &  0  & -2 & -4 \\[1mm]
    12 &  0 & -2  &  0 & 12 \\[1mm]
    -4 & -2 &  0  & -2 &  4 \\[1mm]
     1 & -1 & 1/2 &  1 &  1  
   \end{array} \right),
 \end{equation}

 \noindent where the rows and columns are labeled in the order
 $ i = S $, $ V $, $ T $, $ A $, $ P $.

 As (\ref{fierz}) shows the trace of the Fierz
 crossing matrix is $ Tr(C_{Fierz}) = -1 $.
 This implies that the Fierz crossing matrix has three eigenvalues
 $ \lambda = -1 $ and two eigenvalues $ \lambda = +1 $.
 When discussing physics applications one also needs the corresponding
 Fierz crossing matrix for parity odd products of currents.
 This can easily be derived from multiplying the parity even
 case by $ \gamma_5 $. 

 Amazingly, the charged current interaction of the Standard Model
 $ (V - A)_{\mu} (V - A)^{\mu} $ is an eigenvector of Fierz crossing,
 with eigenvalue $ \lambda = - 1 $.
 When relating the Standard Model and charge retention forms of the
 Lagrangian as in (\ref{chargeretention}) the minus sign from
 Fierz crossing is cancelled from having to commute the Fermion fields
 an odd number of times in order to relate the two forms.

 In pre--QCD days it was considered unfortunate that the eigenvalue of the 
 $ (V - A)_{\mu} (V - A)^{\mu} $ current--current product is negative.
 If it were positive one would have a very natural explanation of the so
 called $ \Delta I = 1/2 $-- or octet rule in weak nonleptonic decays.
 To see this consider the direct products
 $ {\bf 3 \!\otimes\! 3 = \bar{3}_a \oplus 6_s } $ and
 $ {\bf \bar{3} \otimes \bar{3} = 3_a \oplus \bar{6}_s } $.
 With the wrong eigenvalue $ \lambda = + 1 $
 (and the minus sign from Fermi statistics) one would then remain with
 $ {\bf \bar{3}_a \oplus 3_a = 1 \oplus 8} $ and one thus would have explained
 the famous octet rule for nonleptonic weak interactions.
 Nevertheless, with the advent of QCD and colour, it was realized that
 an important class of diagrams in nonleptonic baryon transitions between
 ground state baryons was subject to the octet rule due to the simple
 Fierz property of the nonleptonic current--current product.
 This discovery is sometimes referred to as the Pati--Woo theorem
 \cite{patiwoo} even though the discovery of Pati and Woo was
 predated by the paper \cite{koerner70}.

%%%%%%%%%%%%%%%%%%%%%%%%%%%%%%%%%%%%%%%%%%%%%%%%%%%%%%%%%%%%%%%%%%%%%%%%%%%%%%% 
%
%         Counting Spin Observables 
%
%%%%%%%%%%%%%%%%%%%%%%%%%%%%%%%%%%%%%%%%%%%%%%%%%%%%%%%%%%%%%%%%%%%%%%%%%%%%%%%

\section{Counting Spin Observables}

 When setting up a problem involving the spin of particles it is always
 instructive to first denumerate the complexity of the problem and count
 the number of independent structures of the problem.
 A particularly efficient and physical way to do so is to count the number
 of independent elements of the spin density matrices involved in the
 process. 

 The single spin density matrix $ \rho_{\lambda \lambda'} $ is a
 $ (2 J + 1)(2 J + 1) $ hermitian matrix
 $ (\rho_{\lambda \lambda'} = \rho_{\lambda' \lambda}^{\ast}) $.
 It thus has $ (2 J + 1)^2 $ independent components of which
 $ (J + 1)(2 J + 1) $ are real and $ J (2 J + 1) $ are imaginary.
 The rate is represented by the trace of the density matrix.
 In Table~\ref{Tab2} we list the corresponding degrees of freedom
 for the first few spin cases $ J = 1/2 $, $ 1 $, $ 3/2 $.
 The density matrix is also frequently represented in terms of its
 angular momentum content.
 For the three cases listed in Table~\ref{Tab2} the
 angular momentum content is given by 
 $ 1/2 \otimes 1/2 = 0 \oplus 1 $,
 $ 1 \otimes 1 = 0 \oplus 1 \oplus 2 $ and
 $ 3/2 \otimes 3/2 = 0 \oplus 1 \oplus 2 \oplus 3 $
 where the one--dimensional representation ``$ 0 $''
 is equivalent to the trace or rate.
 It is clear that one obtains the correct respective number
 of spin degrees of freedom $ (2 J + 1)^2 $ in Table~\ref{Tab2} when
 adding up the dimensions of the angular momentum spaces
 appearing in the decomposition
 $ J \otimes J = 0 \oplus \ldots \oplus 2 J $.
 
 \begin{table}
 \caption{Independent components of the single spin density matrix}
   \begin{center}
   \renewcommand{\arraystretch}{1.3}
   \setlength\tabcolsep{5pt}
   \begin{tabular}{ccccc}
     \hline\noalign{\smallskip}
      spin  & rate  & real & imaginary & sum\\
      $ J $ & trace & $ J (2 J + 3) $ & $  J (2 J + 1) $ & $ (2 J + 1)^2 $\\
     \noalign{\smallskip}
     \hline
     \noalign{\smallskip}
     $ J = 1/2 $ & $ 1 $ & $ 2 $ & $ 1 $ & $  4 $\\
     $ J = 1   $ & $ 1 $ & $ 5 $ & $ 3 $ & $  9 $\\
     $ J = 3/2 $ & $ 1 $ & $ 9 $ & $ 6 $ & $ 16 $\\
     \hline
   \end{tabular}
   \end{center}
 \label{Tab2}
 \end{table}

 In the spin $ 1/2 $ case the four spin degrees of freedom
 are the rate, the two real components and the imaginary component
 of the spin $ 1/2 $ density matrix.
 The unnormalized density matrix $ \rho $ is usually written as  

 \begin{eqnarray}
   \label{fermiondensity}
   \rho_{\lambda \lambda'} & = &
   \left( \begin{array}{c@{\quad}c}
     h_{+1/2} h_{1/2}^{\ast} & h_{+1/2} h_{-1/2}^{\ast} \\[2mm]
     h_{-1/2} h_{1/2}^{\ast} & h_{-1/2} h_{-1/2}^{\ast}
   \end{array} \right) \nonumber \\ & = &
   \frac{1}{2} ( \sigma \cdot \1 +
   \vec{\xi} \cdot \vec{\sigma}).
 \end{eqnarray}

 \noindent In the helicity basis the four
 spin degrees of freedom are labelled by

 \vspace*{-3mm}
 \begin{eqnarray}
   \sigma & : & \mbox{ rate }, \nonumber \\
   \xi_x / \sigma & : &
   \mbox{ ``perpendicular'' polarization } P^{\perp}, \nonumber \\ 
   \xi_y / \sigma & : &
   \mbox{ ``normal`` polarization } P^N, \nonumber \\
   \xi_z / \sigma & : &
   \mbox{ ``longitudinal'' polarization } P^l.
 \end{eqnarray}
 \vspace*{-3mm}

 \noindent The $ y $--component $ P^N $ can be seen to be a T--odd measure
 since one needs a vector product to construct the normal to a given plane. 
 For example, in $ \mu $--decay $ P^N $ is determined by the expectation value 
 $ \langle \vec{\sigma}_e \cdot (\vec{\sigma}_\mu \times \vec{p}_e) \rangle $
 which is a $ T $--odd measure.
 In the context of this review $ P^N $ obtains contributions
 only from the imaginary parts of one--loop diagrams.
 For all processes listed in Table~\ref{Tab1} in Sect.~1 the
 one--loop diagrams are real and thus $ P^N = 0 $ in these processes.
 This is different in $ e^+ e^- \rightarrow q \, \bar{q} \, g $
 where the one--loop diagrams possess imaginary parts and
 where one therefore has $ P^N \ne 0 $
 (see e.g. \cite{Groote:1995ky,Kuhn:1985ps}).

 Returning to Table~\ref{Tab2} we shall see that not all $ (2 J + 1)^2 $
 spin degrees of freedom are accessible (angular momentum conservation)
 or measurable (parity) in general.

 In the following we will be concerned with double spin density matrices.
 Their spin degrees of freedom are given by the
 products of the single spin degrees of freedom.
 For the cases discussed in this review the double density matrix is a
 sparsely populated matrix, due to angular momentum conservation and the
 absence of imaginary parts from loop contributions.
 Take, for example, the double spin density matrix of the decay
 $ t (\uparrow) \rightarrow X_b + W^+(\pol) $.
 On naive counting one would expect
 $ N = (1 (\mbox{trace}) + 20 (\mbox{real}) + 15 (\mbox{imaginary})) $
 spin degrees of freedom.
 This number is considerably reduced by considering the following two facts.

 \begin{enumerate}
   \item The one--loop amplitude does not possess an imaginary
   (absorptive) part.
   This can be seen by taking a look at the one--loop vertex corrections
   for the transitions discussed in Table~\ref{Tab1}.
   The vertex correction does not admit of real on--shell intermediate states,
   i.e. it does not have an absorptive part.
   The $ 15 $ imaginary components of the double density matrix are zero.
 
  \item Angular momentum conservation.
   In the rest frame of the top the decay into $ X_b $ and $ W^+ $ is
   back--to--back and thus anti--collinear.
   Since one is summing over the helicities of the $ X_b $ one has
   $ \lambda_{X_b} = \lambda'_{X_b} $ and thus
   $ \lambda_t - \lambda_W = \lambda'_t - \lambda'_W $ (see Fig.~3).
   Taking this constraint into consideration one remains with six diagonal
   and two non--diagonal double spin density matrix elements
   (see Table~\ref{Tab3}) where one refers to diagonal and
   non--diagonal elements when 
   $( \lambda_t, \lambda_W) = (\lambda'_t, \lambda'_W) $ and
   $( \lambda_t, \lambda_W) \ne (\lambda'_t, \lambda'_W) $, respectively.
   One has non--diagonal transitions for the two cases 
   $ \lambda_t - \lambda_W = \pm 1/2 $.
   Concerning the non--diagonal density matrix elements it is easy to
   see from the master formula (\ref{masterformula}) to be discussed
   in Sect.~5 that they generate azimuthal dependences in angular decay
   distributions.  
 \end{enumerate}

 We shall now go through the exercise and count the spin degrees of freedom
 by a different method, namely by counting the number of covariants of
 the process.
 We emphasize that counting spin degrees of freedom via density
 matrix elements is generally safer than counting the number of
 independent covariants because there exist nontrivial identities
 between covariants in $ D = 4 $ dimensions.
 The decay $ t (\uparrow) \rightarrow X_b + W^+(\pol) $
 will illustrate this point.

 Consider the expansion of the spin dependent hadron tensor of the
 process into covariants constructed from the metric tensor,
 the Levi--Civita tensor, the independent momenta of the process and the
 spin four--vector of the top.
 One has

 \vspace{-2mm}
 \begin{eqnarray} %% Tensor-Entwicklung 
   \label{tensor-expansion}
   H^{\mu \nu} & = &  \big( - g^{\mu \nu} \, H_1 +
   p_t^{\mu} p_t^{\nu} \, H_2 - i \epsilon^{\mu \nu \rho \sigma}
   p_{t,\rho} q_{\sigma} \, H_3 \big) \nonumber \\[2mm] & - &
   \big(q \!\cdot\! s_t \big) \big( - g^{\mu \nu} \, G_1 +
   p_t^{\mu} p_t^{\nu} \, G_2 - i \epsilon^{\mu \nu \rho \sigma}
   p_{t,\rho} q_{\sigma} \, G_3 \big) \nonumber \\[2mm] & + &
   \big(s_t^{\mu} p_t^{\nu} + s_t^{\nu} p_t^{\mu} \big) \, G_6 +
   i \epsilon^{\mu \nu \rho \sigma} p_{t \rho} s_{t \sigma} \, G_8 +
   i \epsilon^{\mu \nu \rho \sigma} q_{\rho} s_{t \sigma} \, G_9.
 \end{eqnarray}

 \noindent There are nine covariants and associated with it nine invariants
 which obviously does not agree with the number eight counted before using
 helicity counting.
 The discrepancy arises because there exist a nontrivial four--dimensional
 identity between three of the nine covariants which reads

 \begin{equation} %% Schouten-identity
   \label{schouten-identity}
   q \!\cdot\! s_t \, \epsilon^{\mu \nu \rho \sigma} p_{t,\rho} q_{\sigma} -
   q^2 \epsilon^{\mu \nu \rho \sigma} p_{t,\rho} s_{t \sigma} +
   q \!\cdot\! p_t \, \epsilon^{\mu \nu \rho \sigma} q_{\rho} s_{t,\sigma} = 0.
 \end{equation}

 \noindent This identity can be derived using the Schouten identity which reads

 \begin{table}
   \caption{Non--vanishing double density matrix
    elements in $ t (\pol) \decay X_b + W^+ (\pol) $}
   \begin{center}
   \renewcommand{\arraystretch}{1.5}
   \setlength\tabcolsep{5pt}
   \begin{tabular}{ccc}
     \hline\noalign{\smallskip}
    \hspace{5mm} $ \pp \lambda_t $ \hspace{5mm} &
    \hspace{5mm} $ \pp \lambda_W $ \hspace{5mm} &
    \hspace{5mm} $ \pp \lambda_t - \lambda_W $ \hspace{5mm}\\
    \noalign{\smallskip} \hline \noalign{\smallskip}
     $ \pp 1/2 $ & $ \pp 1 $ & $   - 1/2 $\\
     $ \pp 1/2 $ & $ \pp 0 $ & $ \pp 1/2 $\\
     $ \pp 1/2 $ & $   - 1 $ & $ \pp 3/2 $\\
     $   - 1/2 $ & $ \pp 1 $ & $   - 3/2 $\\
     $   - 1/2 $ & $ \pp 0 $ & $   - 1/2 $\\
     $   - 1/2 $ & $   - 1 $ & $ \pp 1/2 $\\
    \hline
   \end{tabular}
   \end{center}
 \label{Tab3}
 \end{table}

 \begin{equation}
 \label{schouten1}
   T_{\mu [\mu_1 \mu_2 \mu_3 \mu_4 \mu_5]} = 0,
 \end{equation}

 \noindent where the symbol ``$[\ldots]$''
 denotes antisymmetrization and where

 \begin{equation}
   \label{schouten2}
   T_{\mu [\mu_1 \mu_2 \mu_3 \mu_4 \mu_5 ]} =
   g_{\mu \mu_1} \epsilon_{\mu_2 \mu_3 \mu_4 \mu_5} +
   \mbox{cycl.} (\mu_1, \mu_2, \mu_3, \mu_4, \mu_5).
 \end{equation}

 The Schouten identity is just the statement that it is impossible to
 place four index values in an antisymmetric fifth rank tensor,
 or in the language of Young Tableaux, that a Young Tableau with five
 vertical boxes is identically zero in four dimensions.

 This illustration is not only of academic interest but has
 numerical implications since one can get into a terrible mess
 numerically if one works with a redundant set of covariants
 and tries to do matrix inversions involving the overcounted
 set of degrees of freedom.
 Although the Schouten identity seems rather obvious nowadays
 there have been examples in the literature where Schouten--type
 of identities have been overlooked.

 We conclude this section by enumerating the number of
 spin degrees of freedom in the various decay processes
 discussed in this review.
 The processes are listed in Table~\ref{Tab4} together
 with the  number of spin degrees of freedom.
 The corresponding references to the papers in which
 these decays were treated can be found in Table~\ref{Tab1}.
 In this list we have taken into account that the decay
 $ D^{\ast} \rightarrow \pi(\gamma) $ is parity conserving.
 In parantheses we list the number of $ T $--odd observables for
 each process which are zero due to the absence of imaginary
 contributions in the one--loop vertex correction.

 \begin{table}
 \caption{Number of measurable double density matrix
  elements for the processes listed in Table~\ref{Tab1}}
   \begin{center}
   \renewcommand{\arraystretch}{1.3}
   \setlength\tabcolsep{5pt}
   \begin{tabular}{ll}
     \hline\noalign{\smallskip}
     process & spin degrees of freedom \\
     \noalign{\smallskip} \hline \noalign{\smallskip}
      $ t(\pol) \decay b + W^{+}(\pol) $ &
      $ N = 1 + 7 \hspace{2mm} (+ 2) $\\
      $ t(\pol) \longrightarrow b + H^{+} $ &
      $ N = 1 + 1 $ \\
      $ B \phantom{(\pol)} \decay X_c + D_s $ &
      $ N = 1 $ \\
      $ B \phantom{(\pol)} \decay X_c + D_s^{*}(\pol) $ &
      $ N = 1 + 1 $ \\
      $ \Lambda_b(\pol) \decay X_c + D_s \qquad $ &
      $ N = 1 + 1 $ \\
      $ \Lambda_b(\pol) \decay X_c + D_s^{*}(\pol) $ &
      $ N = 1 + 4 \hspace{2mm} (+ 1) $ \\
      $ l(\pol) \decay l'(\pol) + \nu_l + \nu_{l'} $ &
      $ N = 1 + 4 \hspace{2mm} (+ 1) $ \\[1mm]
     \hline
   \end{tabular}
   \end{center}
 \label{Tab4}
 \end{table}

%%%%%%%%%%%%%%%%%%%%%%%%%%%%%%%%%%%%%%%%%%%%%%%%%%%%%%%%%%%%%%%%%%%%%%%%%%%%%%%%
%
%            Angular Decay Distributions 
%
%%%%%%%%%%%%%%%%%%%%%%%%%%%%%%%%%%%%%%%%%%%%%%%%%%%%%%%%%%%%%%%%%%%%%%%%%%%%%%%%

 \section{Angular Decay Distributions}

 It should be clear that one needs to do polarization measurements in order
 to disentangle the full structure of particle interactions.
 Polarization measurements are particularly simple when the particle whose
 polarization one wants to measure decays.
 The angular decay distribution of the decay products reveals information
 on the state of polarization of the decaying particle.
 The information contained in the angular decay distribution
 is maximal when the particle decay is weak.
 The fact that the angular decay distribution reveals information on
 the polarization of the decaying particle is sometimes referred to as
 that the particle decay is self--analyzing.

 There are principally two ways to obtain angular decay distributions
 which we will refer to as the non--covariant and the covariant methods.
 In the non--covariant method one makes use of rotation matrices whereas
 in the covariant method one evaluates scalar products of four--vectors
 involving momenta and spin four--vectors in given reference frames.
 We shall discuss one example each of the two methods.
 
 As an example for the non--covariant method we write down
 the angular decay distribution of polarized top decay
 $ t (\uparrow) \rightarrow b + W^+ $ followed by
 $ W^+ \rightarrow l^+ + \nu_l $.
 The angular decay distribution can
 be obtained from the master formula

 \begin{equation}
   \label{masterformula}
   W (\theta_P, \theta, \phi) \propto
   \hspace{-6mm} \sum\limits_{
   \lambda_W - \lambda^{\prime}_W =
   \lambda_t - \lambda^{\prime}_t} \hspace{-6mm}
   e^{i (\lambda_W - \lambda^{\prime}_W) \phi} \:
   d^1_{\lambda_W 1}(\theta) \: d^1_{\lambda^{\prime}_W 1}(\theta) \:
   H_{\lambda_W \lambda^{\prime}_W}^{\lambda_t \: \lambda^{\prime}_t} \:
   \rho_{\lambda_t \: \lambda^{\prime}_t} (\theta_P),
 \end{equation}

 \noindent where $ \rho_{\lambda_t \: \lambda^{\prime}_t} (\theta_P) $
 is the density matrix of the top quark which reads

 \begin{equation}
   \rho_{\lambda_t \: \lambda^{\prime}_t} (\theta_P) =
   \frac{1}{2} \pmatrix{
   1 + P \cos \theta_P & P \sin \theta_P \vspace{2mm} \cr
   P \sin \theta_P & 1 - P \cos \theta_P }.
 \end{equation}

 \noindent $ P $ is the magnitude of the polarization of the top quark.
 The $ H_{\lambda_W \lambda^{\prime}_W}^{\lambda_t \: \lambda^{\prime}_t} $
 are helicity matrix elements of the hadronic structure function
 $ H_{\mu \nu} $.
 The sum in (\ref{masterformula}) extends over all values of
 $ \lambda_W, \lambda^{\prime}_W, \lambda_t $ and
 $ \lambda^{\prime}_t $ compatible with the constraint
 $ \lambda_W - \lambda^{\prime}_W = \lambda_t - \lambda^{\prime}_t $.
 The second lower index in the small Wigner $ d(\theta) $--function
 $ d^1_{\lambda_W 1} $ is fixed at $ m = 1 $ for zero mass leptons because the
 total $ m $--quantum number of the lepton pair along the $ l^+ $ direction is
 $ m = 1 $.
 Because there exist different conventions for Wigner's
 $ d $--functions we explicate the requisite components that enter
 (\ref{masterformula}): $ d^1_{11} = (1 + \cos\theta) / 2 $,
 $ d^1_{01} = \sin\theta / \sqrt{2} $ and $ d^1_{-11} = (1 - \cos \theta) / 2 $.
 The fact that one has to specify the phase convention of the Wigner's
 $ d $--functions points to one of the weaknesses of the non--covariant method
 to obtain the correct angular decay distributions:
 one has to use one set of consistent phase conventions to obtain
 the correct signs for the angular factors.
 For someone not not so familiar with the angular
 momentum apparatus this is not always simple.
 For the polar angle dependencies the correctness of a
 sign can always be checked by using physics arguments.
 This is more difficult for the azimuthal signs.
   
 Including the appropriate normalization factor the
 four--fold decay distribution is given by \cite{fgklm99,fgkm02}
 
 \begin{eqnarray}
   \label{DiffRate}
   \frac{d \Gamma}{dq_0 d \cos \theta_P d \cos \theta d \phi} & = &
   \frac{1}{4 \pi} \frac{G_F |V_{tb}|^2 m_W^2}{\sqrt{2} \, \pi} \,
   |\vec{q}| \bigg\{ \nonumber \\ & & \hspace{-20mm} +
   \frac{3}{8} (H_U + P \cos \theta_p H_{U^P})
   (1 + \cos^2 \theta) \nonumber \\ & & \hspace{-20mm} +
   \frac{3}{4} (H_L + P \cos \theta_p H_{L^P}) \sin^2 \theta +
   \frac{3}{4} (H_F + P \cos \theta_p H_{F^P}) \cos \theta
   \nonumber \\ & & \hspace{-20mm} +
   \frac{3}{2 \sqrt{2}} P \sin \theta_p H_{I^P} \sin 2 \theta \cos \phi +
   \frac{3}{\sqrt{2}} P \sin \theta_p H_{A^P} \sin \theta \cos \phi \bigg\},
 \end{eqnarray}
 
 \noindent where the helicity structure functions $ H_U $, $ H_L $ etc.
 are linear combinations of the helicity matrix elements
 $ H_{\lambda_W \lambda^{\prime}_W}^{\lambda_t \: \lambda^{\prime}_t} $
 We have taken the freedom to normalize the differential rate
 such that one obtains the total $ t \rightarrow b + W^{+} $ rate upon
 integration {\it and not} the total rate multiplied by the branching
 ratio of the respective $ W^+ $ decay channel.

\begin{figure}[htbp] %% plot IV
   \begin{center} %% def. of angles
     \includegraphics[width=.60 \textwidth]{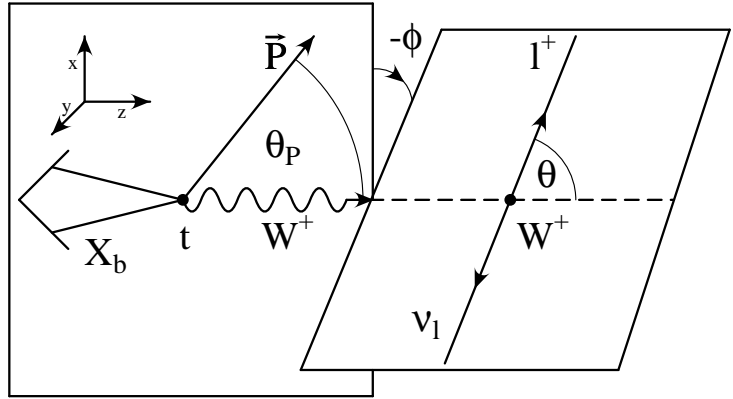}
   \end{center}
   \caption[]{Definition of the angles $ \theta_P $,
    $ \theta $ and $ \phi $ in the cascade decay
    $ t(\pol) \rightarrow X_b + W^+ $ and 
    $ W^+ \rightarrow l^+ + \nu_l $}
   \label{plot4}
 \end{figure}

 The polar angles $ \theta_P $ and $ \theta $, and the
 azimuthal angle $ \phi $ that arise in the full cascade--type
 description of the two--stage decay process
 $ t(\pol) \rightarrow W^{+} (\rightarrow l^{+} + \nu_{l}) + X_{b} $
 are defined in Fig.~\ref{plot4}.
 For better visibility we have oriented the lepton plane with
 a negative azimuthal angle relative to the hadron plane.
 For the hadronic decays of the $ W $ into a pair of light quarks one has
 to replace $ (l^{+}, \nu_{l}) $ by $ (\bar{q}, q) $ in Fig.~\ref{plot4}.
 We mention that we have checked the signs of the angular decay
 distribution (\ref{DiffRate}) using covariant techniques.

 At first sight it seems rather strange that the
 angular analysis is done in two different coordinate systems,
 namely the top quark rest system and the $ W $ rest system.
 This runs counter to common wisdom that invariants
 should always be evaluated in one reference system.
 Some insight into the problem may be gained by using the
 orthonormality and completeness relation of polarization vectors
 to rewrite the contraction of the hadron and lepton tensors
 in a form which exhibits the correctness of using two
 different reference systems for the cascade decay.
 The orthonormality and completeness relation read 

 \begin{eqnarray} %% orthonormality & completeness
   \mbox{Orthonormality:} & \hspace{8mm} &
   g_{\mu \nu} \epsilon^{\ast \mu}(m) \epsilon^{\nu} (m') =
   g_{m m'} \quad m, m' = S, \pm, 0, \\[3mm]
   \mbox{Completeness:} & \hspace{8mm} &
   \sum_{m,m' = S, \pm, 0} \hspace{-6mm}
   \epsilon^{\mu}(m) \epsilon^{\ast \nu}(m') g_{mm'} =
   g^{\mu \nu} \quad \mu, \nu = 0, 1, 2, 3, \qquad 
 \end{eqnarray}

 \noindent where $ g_{mm'} = \mbox{diag} (+,-,-,-) $,
 $ (m, m' = S, \pm, 0) $ and $ g^{\mu \nu} = \mbox{diag}(+,-,-,-) $,
 $ (\mu, \nu = 0, 1, 2, 3) $.
 The scalar or time component of the polarization
 four--vector is denoted by $ S $.
 On using the completeness relation one then has

 \begin{eqnarray}
   \label{lorentztrick}
   L^{\mu\nu}H_{\mu\nu} & = &
   L_{\mu'\nu'} g^{\mu'\mu} g^{\nu'\nu} H_{\mu\nu} =
   L_{\mu'\nu'} \epsilon^{\mu'}(m) \epsilon^{\ast \mu}(m') g_{mm'}
   \epsilon^{\ast \nu'}(n) \epsilon^{\nu}(n')
   g_{nn'} H_{\mu\nu} \nonumber\\[3mm] & = &
   \big( L_{\mu'\nu'} \epsilon^{\mu'}(m) \epsilon^{\ast \nu'}(n) \big) 
   \big( H_{\mu\nu} \epsilon^{*\mu}(m') \epsilon^{\nu}(n') \big) \,
   g_{mm'} g_{nn'}.
 \end{eqnarray}

 \noindent The point is that the two Lorentz contractions appearing on the
 second line of (\ref{lorentztrick}) can be evaluated in two different
 Lorentz frames.
 The leptonic invariant
 $ L_{\mu' \nu'} \epsilon^{\mu'}(m)\epsilon^{\ast \nu'}(n) $
 can be evaluated in the $ (l \nu) $ CM frame
 (or in the $ W^+ $ rest frame) while the hadronic invariant
 $ H_{\mu \nu} \epsilon^{\ast \mu}(m') \epsilon^{\nu}(n') $
 can be evaluated in the rest frame of the top quark.
 Another advantage of this method is that one can
 easily incorporate lepton mass effects as is mandatory for $ b \rightarrow c $
 or the rare $ b \rightarrow s $ transitions if the charged
 lepton or leptons in the final state are $ \tau $ leptons.
 This technique was used to derive angular decay distributions
 including leptonic polarization effects for semileptonic and
 rare bottom meson decays in \cite{Korner:1989qb} and \cite{Faessler:2002ut}.
 Note also that the correct phase choice for the polarization vectors is no
 longer crucial since the polarization vectors always appear as squares in the
 orthonormality and completeness relations.
 
 As an example of the covariant method we discuss the angular
 decay distribution of polarized $ \mu $ decay into a polarized $ e $.
 From helicity counting as described in Sect.~4 one knows that there are
 altogether five spin--dependent structure functions and one
 spin--independent structure function describing the leptonic decay
 of a polarized muon into a polarized electron.
 We thus define a spin--dependent differential rate in terms of
 six invariant structure functions $ A_i $.
 Accordingly one has \cite{fgkm03electron} 

 \begin{eqnarray} %% spin-dependent rate in terms of invariants
   \label{invrate}
   \frac{d \Gamma}{dx \, d \! \cos \theta_P } & = & A_1 +
   \frac{1}{m_{\mu}} A_2 (p_{e} \!\cdot\! s_{\mu}) +
   \frac{1}{m_{\mu}} A_3 (p_{\mu} \!\cdot\! s_{e}) +
   \frac{1}{m_{\mu}^2} A_4 (p_e \!\cdot\! s_{\mu}) \,
   (p_{\mu} \!\cdot\! s_{e}) \nonumber \\[2mm] & + &
   A_5 (s_{\mu} \!\cdot\! s_{e}) +
   \frac{1}{m_{\mu}^2} A_6 \,
   \epsilon_{\alpha \beta \gamma \delta} \,
   p_{\mu}^{\alpha} \, p_{e}^{\beta} \,
   s_{\mu}^{\gamma} \, s_{e}^{\delta}.
 \end{eqnarray}

\begin{figure}[thpb] %% plot V
   \begin{center} %% def. of angles
     \includegraphics[width=.30 \textwidth, clip=]{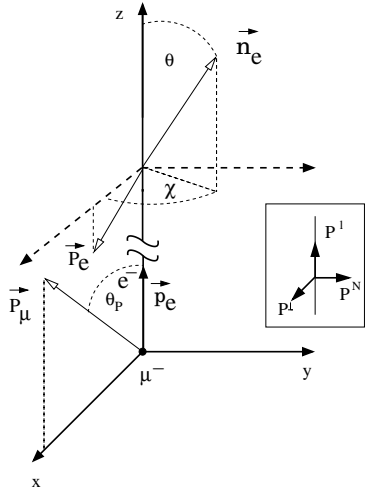}
   \end{center}
   \vspace*{-3mm}
   \caption[]{Definition of the angles $ \theta_P $,
    $ \theta $ and $ \chi $ in polarized muon decay}
   \label{plot5}
 \end{figure}

 Eq.~(\ref{invrate}) will be evaluated in the rest system
 of the muon where $ p_{\mu} = (m_{\mu}; 0, 0, 0) $ and 
 $ p_e = (E_e; 0, 0, |\vec{p}_e|) = $
 $ (m_{\mu}/2) (x; 0, 0, x \beta) $.
 The velocity of the electron is denoted by
 $ \beta = \sqrt{1 - 4 y^2 / x^2} $ where
 $ y = m_e / m_{\mu} $ and $ x = 2 E_e / m_{\mu} $
 denotes the scaled energy of the electron.
 In the rest frame of the $ \mu^- $ the polarization
 four--vectors of the $ \mu^- $ and $ e^- $ are given by

 \begin{eqnarray} % *** polarization vectors *** %
   \label{polarizationvectors}
   s_{\mu}^{\alpha} & = &
   (0; \vec{\zeta}_{\mu}), \\[3mm]
   s_{e}^{\alpha} & = &
   (\frac{\vec{n}_e \!\cdot\! \vec{p}_e}{m_e};
   \vec{n}_{e} +
   \frac{\vec{n}_e \!\cdot\! \vec{p}_e}{m_e (E_e + m_e)}
   \vec{p}_e),
 \end{eqnarray}
 
 \noindent where the polarization three--vector
 $ \vec{\zeta}_{\mu} $ of the $ \mu^- $
 and the quantization axis $ \vec{n}_e $ of
 the spin of the $ e^- $ in their respective
 rest frames read (see Fig.~\ref{plot5})
 
 \begin{equation} % *** polarization three-vector polmu *** %
   \label{restframepolmu}
   \vec{\zeta}_{\mu} = (\sin \theta_P, 0, \cos \theta_{P})  
 \end{equation}
 
 \noindent and
 
 \begin{equation} % *** polarization three-vector polelectron *** %
   \label{restframepolelectron}
   \vec{n}_{e} =
   (\sin \theta \cos \chi,
    \sin \theta \sin \chi,
    \cos \theta).  
 \end{equation}

 Eq.~(\ref{restframepolmu}) holds for $ 100 \% $ polarized muons.
 For partially polarized muons with magnitude of polarization $ P $ the
 representation (\ref{restframepolmu}) has to be multiplied by $ P $ such
 that $ \vec{P}_{\mu} = P \vec{\zeta}_{\mu} $.
 
 The scalar products in (\ref{invrate}) can then be evaluated with the result

 \begin{eqnarray}
   \hspace{16mm} p_e \!\cdot\! s_\mu & = & -
   \frac{m_{\mu}}{2} x \beta P \cos \theta_P, \nonumber \\
   \hspace{16mm} p_{\mu} \!\cdot\! s_e & = & \pp
   \frac{m_{\mu}}{2 y} x \beta \cos \theta, \nonumber \\
   \hspace{16mm} s_{\mu} \!\cdot\! s_e & = & -
   P \sin \theta_P \sin \theta \cos \chi, \nonumber \\
   \epsilon_{\alpha \beta \gamma \delta} \,
   p_{\mu}^{\alpha} \, p_e^{\beta} \,
   s_{\mu}^{\gamma} \, s_e^{\delta} & = & \pp
   \frac{m^2_\mu}{2} x \beta 
   P \sin \theta_P \sin \theta \sin \chi.
 \end{eqnarray}

 \noindent Including the correct normalization one finally arrives
 at the angular decay distribution \cite{fgkm03electron} 
 
 \begin{eqnarray}   % ***   spin-dependent  *** %
   \label{diffrate} % *** differential rate *** %
   \frac{d \Gamma}{dx \, d \! \cos \theta_P } & = &
   \beta x \, \Gamma_0 (G_1 + G_2 P \cos \theta_P +
    G_3 \cos \theta + G_4 P \cos \theta_P \cos \theta
   \nonumber \\[3mm] & + &
    G_5 P \sin \theta_P \sin \theta \cos \chi +
    G_6 P \sin \theta_P \sin \theta \sin \chi ),
 \end{eqnarray}
 
 \noindent where the relation between the invariant structure functions
 $ A_i $ and the frame--dependent spectrum functions
 $ G_i $ can be found in \cite{fgkm03electron}.
 $ G_1 $ is the unpolarized spectrum function,
 $ G_2 $ and $ G_3 $ are single spin polarized spectrum functions
 referring to the spins of the $ \mu^- $ and $ e^- $, resp.,
 and $ G_4 $, $ G_5 $ and $ G_6 $ describe
 spin--spin correlations between the spin vectors
 of the muon and electron.
 $ G_6 $ represents a so--called $ T $--odd observable.
 This is evident when rewriting the angular factor
 multiplying $ G_6 $ in (\ref{diffrate}) in triple--product
 form, i.e. $ \sin \theta_P \sin \theta \sin \chi = $
 $ |\vec{p}_e|^{-1} \vec{p}_e \cdot (\vec{\zeta}_\mu \times \vec{n}_e) $.
 As stated before $ G_6 $ is identically zero in the Standard Model since,
 on the one hand, the weak coupling constant $ G_F $ is real and,
 on the other hand, the loop contributions do not generate imaginary parts.
 In the terminology of Sect.~4 $ G_1 - G_4 $ are diagonal structure
 functions, while the azimuthally dependent structure functions
 $ G_5 $ and $ G_6 $ are non--diagonal.
 As mentioned before, the structure function $ G_5 $ is proportional
 to the electron mass and therefore vanishes for $ m_e \rightarrow 0 $.
 Naturally, the angular decay distribution (\ref{diffrate}) can also be
 derived in the helicity formalism. 

 In the two examples discussed in this section we chose to orient the z--axis
 along the three--momentum of one of the particles.
 In this case one says that the angular analysis is done in the helicity system.
 Of course, other choices of the orientation of the z--axis are possible.
 A popular choice is the transversity system where the z--axis is normal to
 a given plane spanned by the momenta of the decay products.
 A more detailed discussion of the choice of frames can be found in
 \cite{Korner:1989qb} where the angular decay distribution in the
 semileptonic decays $ B \rightarrow (D, D^{\ast}) + l + \nu $
 including lepton mass effects was investigated.
 We mention that the transversity system is the preferred choice
 when discussing the {\sl CP} properties of the final state
 \cite{Dunietz:1990cj}.

%%%%%%%%%%%%%%%%%%%%%%%%%%%%%%%%%%%%%%%%%%%%%%%%%%%%%%%%%%%%%%%%%%%%%%%%%%%%%%%%
%
%           One-Loop Amplitudes
%
%%%%%%%%%%%%%%%%%%%%%%%%%%%%%%%%%%%%%%%%%%%%%%%%%%%%%%%%%%%%%%%%%%%%%%%%%%%%%%%%

\section{One-Loop Amplitude}

 We present our results in terms of the three vector current amplitudes
 $ F_{i}^{V} $ $ (i=1,2,3) $ and the three axial vector current amplitudes
 $ F_{i}^{A} $ $ (i=1,2,3) $ defined by (
 $ J^V_\mu = \bar{q}_b \gamma_\mu q_t $, 
 $ J^A_\mu = \bar{q}_b \gamma_\mu \gamma_5 q_t $) 

 \alpheqn
 \begin{eqnarray}
   \label{formfactor} %% definition of formfactors
   \langle b(p_b) | J^V_{\mu} | t(p_t) \rangle & = &
   \bar{u}_b(p_b) \Big\{ \gamma_{\mu} F_1^V + p_{t, \mu} F_2^V +
   p_{b,\mu} F_3^V \Big\} u_t(p_t), \\[2mm] 
   \langle b(p_b) |J^A_{\mu} | t(p_t) \rangle & = &
   \bar{u}_b(p_b) \Big\{ \gamma_{\mu} F_1^A + p_{t, \mu} F_2^A +
   p_{b,\mu} F_3^A \Big\} \gamma_5 u_t(p_t).
 \end{eqnarray}  
 \reseteqn

 \noindent In this Section we choose to label the current transition
 according to the transition $ t \rightarrow b $ as in \cite{fgkm02}.
 As emphasized before, up to the colour factor identical expressions
 are obtained for the $ \mu \rightarrow e $ transition if the
 leptonic four--fermion interaction is written in the charge retention form.
 For the form factors one obtains \cite{fgkm02}
 
 \alpheqn
 \begin{eqnarray}
   \label{oneloop}
   F_1^V & = & 1 + \frac{\alpha_s}{4 \pi} C_F \bigg\{ -
   \frac{m_t^2 + m_b^2 - q^2}{m_t^2 \sqrt{\lambda}} \bigg[
   2 \, \Li (1 - w_1^2) - 2 \, \Li \bigg( 1 - \frac{w_1}{w_{\mu}}
   \bigg) \\[2mm] & + &
   \frac{1}{2} \ln \bigg( \frac{\Lambda^4}{m_b^2 m_t^2} \bigg)
   \ln (w_1 w_{\mu}) + \ln \bigg( \frac{w_1^3}{w_{\mu}} \bigg)
   \ln \bigg( \frac{w_{\mu} (1 - w_1^2)}{w_{\mu} - w_1} \bigg) \bigg] -
   \ln \bigg( \frac{\Lambda^4}{m_b^2 m_t^2} \bigg)
   \nonumber \\[2mm] & - &
   \frac{m_t^2 - m_b^2}{2 q^2} \ln \bigg( \frac{m_b^2}{m_t^2} \bigg) -
   4 + \ln (w_1 w_{\mu}) \bigg( \frac{m_t^2 \sqrt{\lambda}}{2 q^2} -
   \frac{(m_t + m_b)^2 - q^2}{m_t^2 \sqrt{\lambda}} \bigg)
   \bigg\}, \nonumber \\[2mm]
   F_2^V & = & \frac{\alpha_s}{4 \pi} C_F
   \frac{m_t - m_b}{q^2} \bigg\{
   2 - \bigg( \frac{m_t + 2 m_b}{m_t - m_b} -
   \frac{m_t^2 - m_b^2}{q^2} \bigg)
   \ln \bigg( \frac{m_b^2}{m_t^2} \bigg) \\[2mm] & - &
   \bigg( \frac{m_t^2 \sqrt{\lambda}}{q^2} - \frac{m_b}{m_t - m_b}
   \frac{q^2 + (m_t - m_b)(3 m_t + m_b)}{m_t^2 \sqrt{\lambda}} \bigg)
   \ln (w_1 w_{\mu}) \bigg\}, \nonumber \\[2mm]
   F_3^V & = & F_3^V (m_t, m_b) = F_2^V (m_b, m_t),
 \end{eqnarray}
 \reseteqn

 \noindent where we have denoted the scaled (small) gluon mass by
 $ \Lambda = m_g / m_t $.
 We define $ \lambda = 1 + x^4 + y^4 - 2x^2 y^2 - 2x^2 - 2y^2 $
 ($ x = q^2/m_t^2 $, $ y = m_b/m_t $) and use the abbreviations
 
 \begin{equation} % *** Def. of w_1,mu *** %
   \label{w1wmu}
   w_1 = \frac{x}{y} \!\cdot\!
   \frac{1 - x^2 + y^2 - \sqrt{\lambda}}
   {1 + x^2 - y^2 + \sqrt{\lambda}}, \qquad
   w_{\mu} = \frac{x}{y} \!\cdot\!
   \frac{1 - x^2 + y^2 - \sqrt{\lambda}}
   {1 + x^2 - y^2 - \sqrt{\lambda}}.
 \end{equation}

 The axial vector amplitudes $ F_i^A $ can be obtained from the vector 
 amplitudes by the replacement $ m_t \rightarrow - m_t $, i.e. one has
 $ F_i^A(m_t) = F_i^V(- m_t) $ $ (i = 1, 2, 3) $.
 Our one--loop amplitudes are linearly related to the one--loop amplitudes
 given in \cite{gounaris} after correcting for a typo in \cite{gounaris}
 (see also \cite{schilcher81}).

 Note that the infrared singularities proportional to $ \ln \Lambda $
 and the would--be mass singularities (also called collinear singularities)
 proportional to $ \ln m_b $ all reside in the Born term form factors
 $ F_1^V $ and $ F_1^A $.
 They are eventually cancelled by the corresponding
 singularities in the tree graph contribution.
 A look at the arguments of the log and dilog functions
 in \ref{oneloop} shows that the one--loop contribution
 is purely real as remarked on earlier. 

 In the case of the electroweak radiative corrections to
 $ t \rightarrow b + W^+ $ there are altogether $ 18 $ different
 vertex correction diagrams in the Feynman--'t Hooft gauge \cite{dgkm03}
 as compared to the one vertex correction diagram in the QCD
 and the $ \mu $--decay cases discussed in this section.
 In addition to the massive one--loop three--point functions one has to 
 calculate the many massive one--loop two--point functions needed in
 the renormalization program.
 We have recalculated all one--loop contributions analytically and
 have checked them analytically and numerically with the help of a
 XLOOPS/GiNaC package that automatically calculates one--loop two--point
 and three--point functions \cite{Bauer:2001ig}.
 Our one--loop results agree with the results of \cite{Denner:1990ns}.
 The results are too lengthy to be reproduced here in analytical form.
 The full analytical results will be given in a
 forthcoming publication \cite{dgkm04}.

%%%%%%%%%%%%%%%%%%%%%%%%%%%%%%%%%%%%%%%%%%%%%%%%%%%%%%%%%%%%%%%%%%%%%%%%%%%%%%%%
%
%         Tree Graph Amplitudes 
%
%%%%%%%%%%%%%%%%%%%%%%%%%%%%%%%%%%%%%%%%%%%%%%%%%%%%%%%%%%%%%%%%%%%%%%%%%%%%%%%%

\section{Tree-Graph Contribution}

 As emphasized earlier on, the \NLO tree graph contributions for the
 QED and QCD cases are identical up to \NLO
 except for a trivial colour factor in the case of QCD.
 Differences set in only at \NNLO where three--gluon
 coupling contributions come in.
 We choose to present and discuss the \NLO tree graph contributions
 in the QED radiative corrections to leptonic $ \mu $--decay as
 written down in \cite{fgkm03electron}.

 We begin with the Born term contribution.
 This is an exercise that everyone has probably gone through
 before in the unpolarized case.
 The point that not everyone is familiar with is
 that it is very simple to include the spin of fermions at the Born term level.
 All that is needed is the substitution $ p \rightarrow \bar{p} $
 for the fermion's momenta where the notation $ \bar{p} $ is explained
 in the following.
 The $ p \rightarrow \bar{p} $ rule is best explained by looking at the
 trace expression of the charge--side tensor which reads
  
 \begin{equation} % *** charged Born term *** %
   \label{borntrace}
   C_{Born}^{\alpha \beta} =
   \frac{1}{4} \mbox{Tr} \Big\{
   (\slp_{e} + m_{e})
   (\1 + \gamma_5 \sls_{e})
   \gamma^{\alpha}
   (\1 - \gamma_5)
   (\slp_{\mu} + m_{\mu})
   (\1 + \gamma_5 \sls_{\mu})
   \gamma^{\beta}
   (\1 - \gamma_5) \Big\}.
 \end{equation}

 \noindent The dependence on the polarization four--vectors of the
 $ \mu^- $ and $ e^- $ has been retained in (\ref{borntrace}).

 Since only even--numbered $ \gamma $--matrix strings survive between the
 two $ (\1 - \gamma_5) $--factors in (\ref{borntrace}) one can compactly
 write the result of the trace evaluation as

 \begin{equation} % *** trace of Born term *** %
   \label{bornspur}
   C_{Born}^{\alpha \beta} = 2 (
   \bar{p}_{\mu}^{\beta} \, \bar{p}_{e}^{\alpha} +
   \bar{p}_{\mu}^{\alpha} \, \bar{p}_{e}^{\beta} -
   g^{\alpha \beta} \, \bar{p}_{\mu} \!\cdot\! \bar{p}_{e} +
   i \epsilon^{\alpha \beta \gamma \delta}
   \bar{p}_{e,\gamma} \bar{p}_{\mu,\delta}),
 \end{equation}

 \noindent where

 \alpheqn
 \begin{eqnarray} % *** abbreviations *** %
   \bar{p}_{\mu}^{\alpha} & = &
      p_{\mu}^{\alpha} -
      m_{\mu} s_{\mu}^{\alpha}, \\[2mm]
   \bar{p}_{e}^{\alpha} & = &
      p_{e}^{\alpha} -
      m_{e} s_{e}^{\alpha},
 \end{eqnarray}
 \reseteqn

 \noindent and where $ s_{\mu}^{\alpha} $ and $ s_{e}^{\alpha} $ are the
 polarization four--vectors of the $ \mu^- $ and $ e^- $.

 In the QED case one has the simplifying feature that the contribution of the
 last term antisymmetric in $ (\alpha \beta ) $ can be dropped.
 The reason is that the dependence on the momentum directions of the
 $ \bar{\nu}_e $-- and $ \nu_{\mu} $--neutrinos is completely integrated out
 in the differential rate.
 Thus the neutrino--side of the interaction can only depend on the spatial
 piece of the second rank tensor build from the momentum transfer to the
 neutrinos (the neutrinos are treated as massless) which is symmetric in  
 $ (\alpha \beta) $.
 Upon contraction with the charge--side tensor
 the antisymmetric piece drops out.

 Unfortunately the  $ p \rightarrow \bar{p} $ trick
 no longer works at \NLO order.
 But still, by using the $ \bar{p} $--notation, the result
 can be presented in a very compact form.
 The result for the \NLO $ \mu \rightarrow e $
 tree graph contribution reads

 \begin{eqnarray} %% matrix element
 \label{matrix-element}
   C^{\alpha \beta} & = &
   \sum_{\gamma \mathrm{-spin}}
   {\cal M}^{\alpha} {\cal M}^{\beta \dagger} =
   \frac{e^2}{2} \bigg\{ \bigg(
   \frac{k \!\cdot\! \bar{p}_e - m_e^2}
   {k \!\cdot\! p_e} +
   \frac{p_{\mu} \!\cdot\! \bar{p}_e}
   {k \!\cdot\! p_{\mu}} \bigg)
   \frac{k^{\alpha} \bar{p}_{\mu}^{\beta} +
   k^{\beta} \bar{p}_{\mu}^{\alpha} -
   k \!\cdot\! \bar{p}_{\mu} g^{\alpha \beta}}
   {k \!\cdot\! p_e} \nonumber \\[1mm] & + & \bigg(
   \frac{k \!\cdot\! \bar{p}_{\mu} + m_{\mu}^2}
   {k \!\cdot\! p_{\mu}} -
   \frac{p_e \!\cdot\! \bar{p}_{\mu}}
   {k \!\cdot\! p_e} \bigg)
   \frac{k^{\alpha} \bar{p}_e^{\beta} +
   k^{\beta} \bar{p}_e^{\alpha} -
   k \!\cdot\! \bar{p}_e g^{\alpha \beta}}
   {k \!\cdot\! p_{\mu}}
   \nonumber \\[1mm] & + &
   (k \!\cdot\! \bar{p}_{\mu})
   \frac{p_e^{\alpha} \bar{p}_{e}^{\beta} +
   p_e^{\beta} \bar{p}_{e}^{\alpha} -
   m_e^2 g^{\alpha \beta}}
   {(k \!\cdot\! p_e)(k \!\cdot\! p_{\mu})} -
   (k \!\cdot\! \bar{p}_e)
   \frac{p_{\mu}^{\alpha} \bar{p}_{\mu}^{\beta} +
   p_{\mu}^{\beta} \bar{p}_{\mu}^{\alpha} -
   m_{\mu}^2 g^{\alpha \beta}}
   {(k \!\cdot\! p_e)(k \!\cdot\! p_{\mu})}
   \nonumber \\[1mm] & + &
   (k \!\cdot\! \bar{p}_e)
   \frac{p_e^{\alpha} \bar{p}_{\mu}^{\beta} +
   p_e^{\beta} \bar{p}_{\mu}^{\alpha} -
   p_e \!\cdot\! \bar{p}_{\mu} g^{\alpha \beta}}
   {(k \!\cdot\! p_e)^2} -
   (k \!\cdot\! \bar{p}_{\mu})
   \frac{p_{\mu}^{\alpha} \bar{p}_e^{\beta} +
   p_{\mu}^{\beta} \bar{p}_e^{\alpha} -
   p_{\mu} \!\cdot\! \bar{p}_e g^{\alpha \beta}}
   {(k \!\cdot\! p_{\mu})^2} \bigg\}
   \nonumber \\[1mm] & - &
   \frac{e^2}{2}
   \bigg( \frac{m_{\mu}^2}{(k \!\cdot\! p_{\mu})^2} \!+\!
   \frac{m_e^2}{(k \!\cdot\! p_{e})^2} \!-\!
   \frac{2 p_e \!\cdot\! p_{\mu}}
   {(k \!\cdot\! p_e) (k \!\cdot\! p_{\mu})} \bigg)
   (\bar{p}_e^{\alpha} \bar{p}_{\mu}^{\beta} +
   \bar{p}_e^{\beta} \bar{p}_{\mu}^{\alpha} -
   \bar{p}_e \!\cdot\! \bar{p}_{\mu} g^{\alpha \beta}).
 \end{eqnarray}

 \noindent The momentum of the radiated photon is denoted by $ k $.
 For the aforementioned reason we have dropped the contribution of 
 antisymmetric terms.
 When the antisymmetric $ \epsilon $--tensor pieces are kept, and
 when one replaces $ e \rightarrow g_s $ and $ 1 \rightarrow N_c C_F = 4 $
 to account for colour, one recovers the \NLO QCD corrected hadronic tensor
 for the $ t \rightarrow b $ transition listed in \cite{fgkm02}.  

 In the last line of (\ref{matrix-element}) we have isolated the
 infrared singular piece of the charge--side tensor which is given by the
 usual soft photon factor multiplying the Born term contribution.
 Technically this is done by writing

 \begin{equation}
  C^{ (\alpha) \alpha \beta} = \bigg( C^{ (\alpha) \alpha \beta} - 
  C^{ (\alpha) \alpha \beta}(\mathrm{soft photon}) \bigg) +
  C^{ (\alpha) \alpha \beta}(\mathrm{soft photon}).
 \end{equation}

 \noindent The remaining part of the charge--side tensor in
 (\ref{matrix-element}) is referred to as the hard photon contribution.
 It is infrared finite and can thus be integrated without a
 regulator photon mass.
 
 Analytic phase space integrations without a
 regulator photon mass are much simpler.
 The regulator photon mass would introduce a new mass scale
 into the problem which, as emphasized before, would complicate
 the phase space integrations.

 In the phase space integration over the photon momentum the infrared
 singular piece is regularized by introducing a (small) photon mass
 resulting in a logarithmic mass divergence in the photon mass.
 Since the integrand of the soft photon piece is much simpler the
 phase space integration can be done analytically by carefully taking the
 appropriate mass zero limits in the integrations.
 The infrared divergence shows up as a logarithmic mass divergence
 in the photon mass.
 In addition, since the soft photon piece factors the Born term tensor,
 the Born term tensor can be pulled out of the phase space integrations.
 The resulting singular soft photon piece is therefore universal in the
 sense that it is the same for all spin structure functions,
 i.e. once the soft photon integration has been done for the
 rate the work is done.
 Eventually the infrared singular piece is cancelled by the
 corresponding singular piece in the one--loop contributions.
 
 The approximation where only the soft photon (or gluon) piece is retained 
 in the \NLO tree--graph hadron tensor is called the soft photon (or gluon)
 approximation.
 From what has been said before it is clear that a \NLO radiative
 correction calculation is much simplified in the soft photon (or gluon)
 approximation.
 In the next section we shall take a specific example, namely
 the process $ e^+ e^- \rightarrow t \, \bar{t} \, (g) $, 
 to investigate the quality of the soft gluon approximation.

 \begin{figure}[htbp] %% plot VI
   \begin{center} %% def. of angles
     \includegraphics[width=.24 \textwidth]{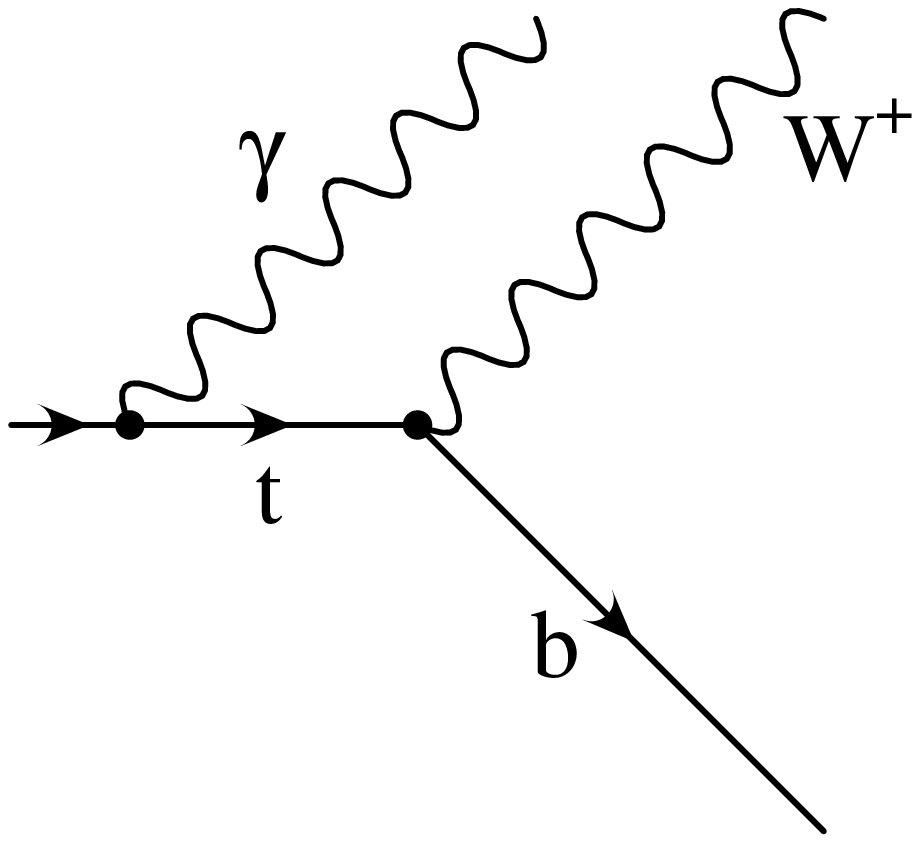}
     \includegraphics[width=.24 \textwidth]{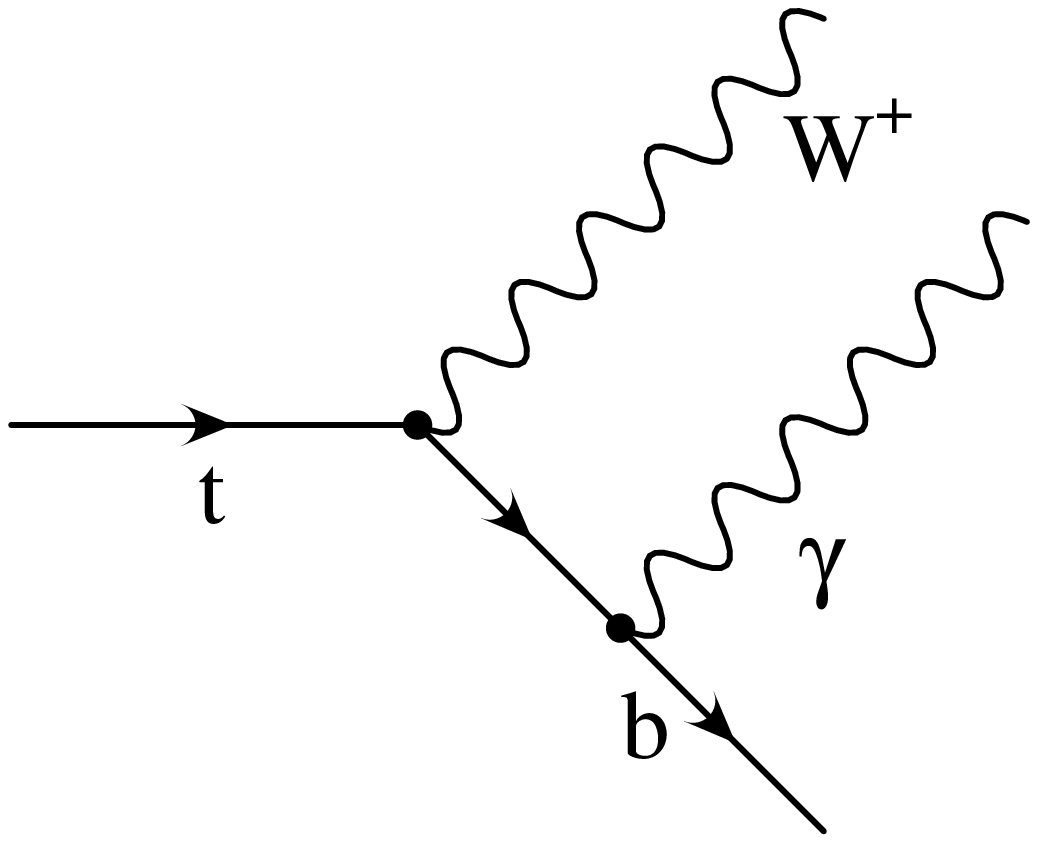}
     \includegraphics[width=.24 \textwidth]{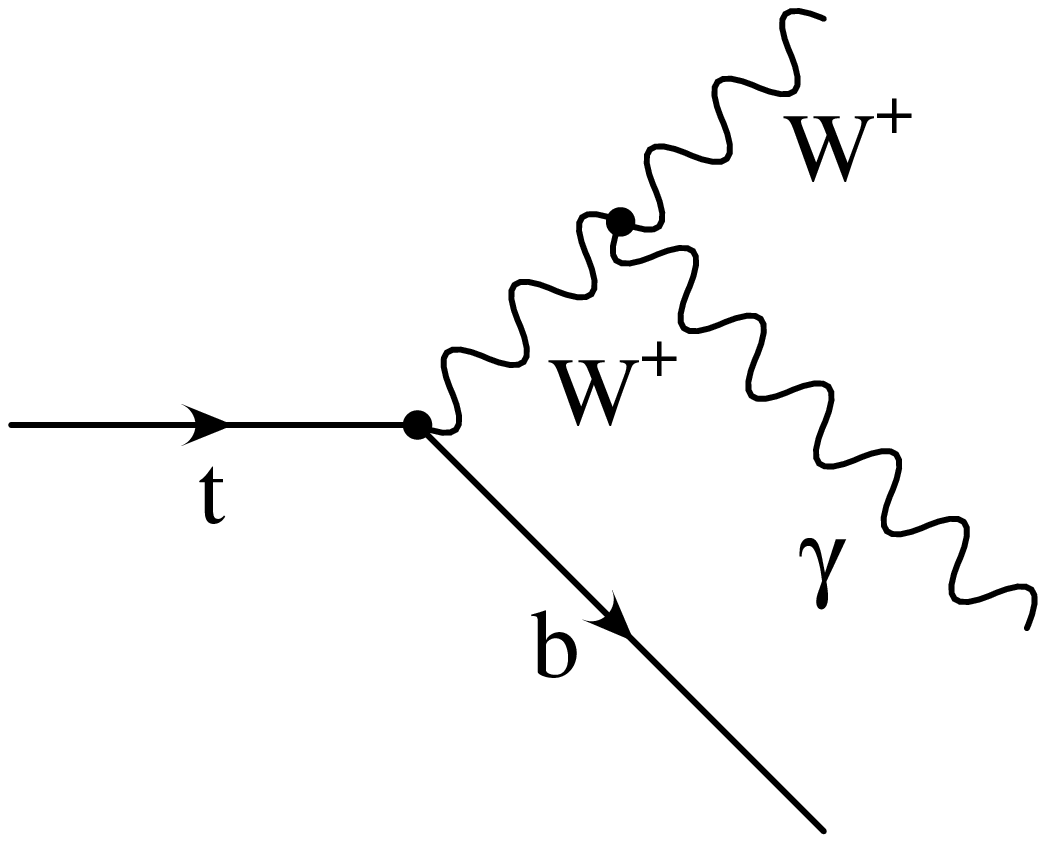}
     \includegraphics[width=.24 \textwidth]{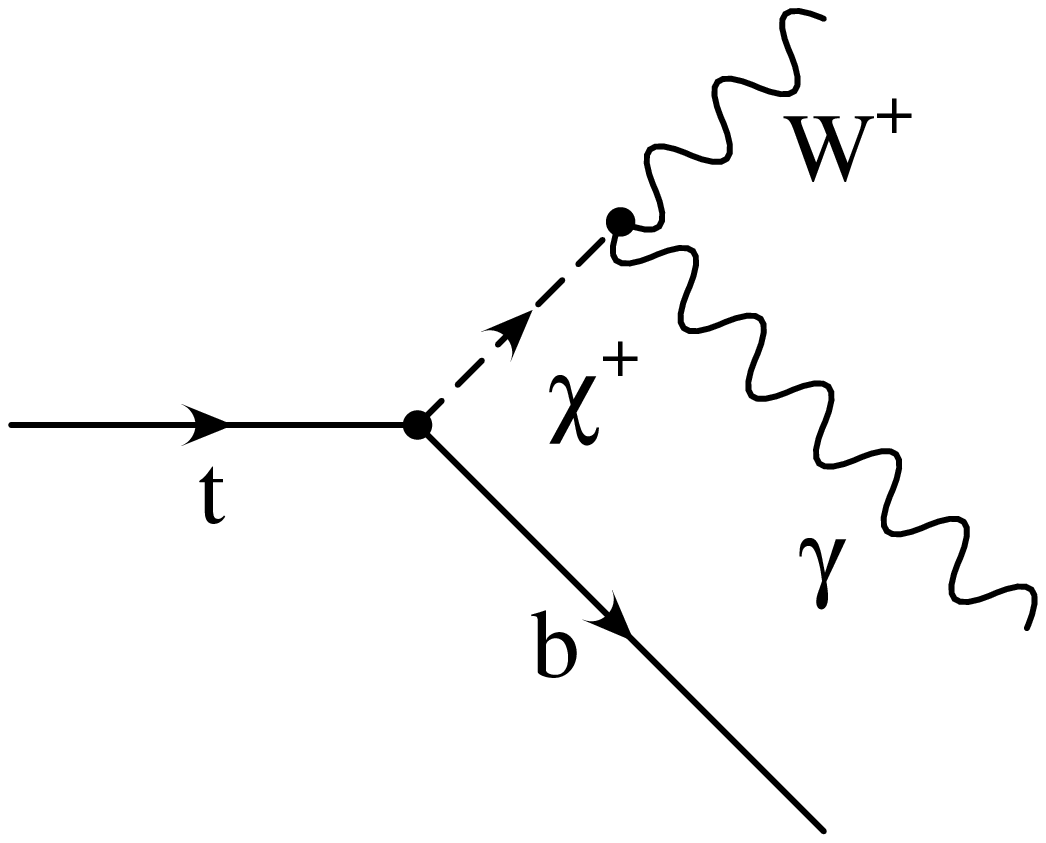}
   \end{center}
   \caption[]{Tree--level Feynman graphs for
    $ t \rightarrow b + W^+ + \gamma $ in
    the Feynman--'t Hooft gauge}
   \label{plot6}
 \end{figure}

 It is interesting to note that the \NLO hadronic tensor in the
 elektroweak corrections to the  process $ t \rightarrow b + W^+ $
 is only marginally more complicated than the corresponding
 tree graph contribution (\ref{matrix-element}) although the
 number of contributing diagrams has doubled to four \cite{dgkm03}
 (see Fig.~\ref{plot6}).
 We do not list the finite piece of the hadronic tensor in this review but
 only write down the soft photon factor which multiplies the Born term tensor
 just as in (\ref{matrix-element}).
 When discussing the decay $ t \rightarrow b + W^+ $ one has to of
 course include the antisymmetric $ \epsilon $--tensor piece in the
 Born term tensor.
 The soft photon contribution reads

 \begin{eqnarray} %% elw. soft photon facto
   \label{softphotonfactor}
   - g^{\mu \nu} A_\mu^{(\mathrm{s.ph.})}
   A_\nu^{(\mathrm{s.ph.})} & = & - e^2 \bigg(
   \frac{Q_t^2 \, m_t^2}{(p_t \!\cdot\! k)^2} +
   \frac{Q_b^2 \, m_b^2}{(p_b \!\cdot\! k)^2} +
   \frac{Q_W^2 \, m_W^2}{(q \!\cdot\! k)^2} \nonumber \\[2mm] & - &
   \frac{2 Q_t Q_b \, p_t \!\cdot\! p_b}
   {(p_t \!\cdot\! k) (p_b \!\cdot\! k)} -
   \frac{2 Q_t Q_W \, p_t \!\cdot\! q}
   {(p_t \!\cdot\! k) (q \!\cdot\! k)} +
   \frac{2 Q_b Q_W \, p_b \!\cdot\! q}
   {(p_b \!\cdot\! k) (q \!\cdot\! k)} \bigg),
 \end{eqnarray}
 
 \noindent where the soft photon amplitude is given by
 (see e.g. \cite{weinberg1})
 
 \begin{equation} %% soft photon amplitude
 \label{softphotonamp}
   A^{(\mathrm{s.ph.}) \mu} \epsilon^{\ast}_\mu =
   e \bigg( \frac{Q_t \, p_t^\mu}{p_t \!\cdot\! k} - 
   \frac{Q_b \, p_b^\mu}{p_b \!\cdot\! k} -
   \frac{Q_W \, q^\mu}{q \!\cdot\! k} \bigg) \epsilon^{\ast}_\mu. 
 \end{equation}

 \noindent $ Q_t = 2/3 $, $ Q_b = -1/3 $ and $ Q_W = 1 $ are the
 electric charges of the top quark, the bottom quark and the $ W $--boson,
 resp., in units of the elementary charge $ e $.
 Gauge invariance of the soft--photon amplitude is easily verified when
 replacing $ \epsilon^{\ast}_\mu \rightarrow k_\mu $ in (\ref{softphotonamp}).
 It is then just a statement about charge conservation
 $ Q_t - Q_b - Q_W = 0 $.
 It is noteworthy that by setting $ Q_t = Q_b = 1 $ and $ Q_W = 0 $
 in (\ref{softphotonfactor}), one recovers the soft photon contribution in
 (\ref{matrix-element}).
 In fact, the whole \NLO electroweak tree--graph tensor in \cite{dgkm03}
 reduces to the corresponding \NLO QED tensor with the above charge
 replacements.
 In the same vein, the replacements $ e \rightarrow g_s $ and 
 $ 1 \rightarrow N_c C_F = 4 $, to account for colour, and the
 replacements of the charge factors by $ Q_t = Q_b = 1 $ and $ Q_W = 0 $ will
 bring one from the electroweak case to the QCD case.

%%%%%%%%%%%%%%%%%%%%%%%%%%%%%%%%%%%%%%%%%%%%%%%%%%%%%%%%%%%%%%%%%%%%%%%%%%%%%%%
%
% NLO radiative corrections to e+ e- -> t \bar{t} (g)
% in the soft gluon approximation
%
%%%%%%%%%%%%%%%%%%%%%%%%%%%%%%%%%%%%%%%%%%%%%%%%%%%%%%%%%%%%%%%%%%%%%%%%%%%%%%%

 \section{{\boldmath $ N\!LO $} Radiative Corrections to \pf
  $ e^+ e^- \rightarrow t \, \bar{t} \, (g) $ in the Soft Gluon Approximation}
 
 \begin{figure}[htbp] %% plot VII
   \begin{center} %% Fig. of S. Groote
     \includegraphics[width=.70 \textwidth]{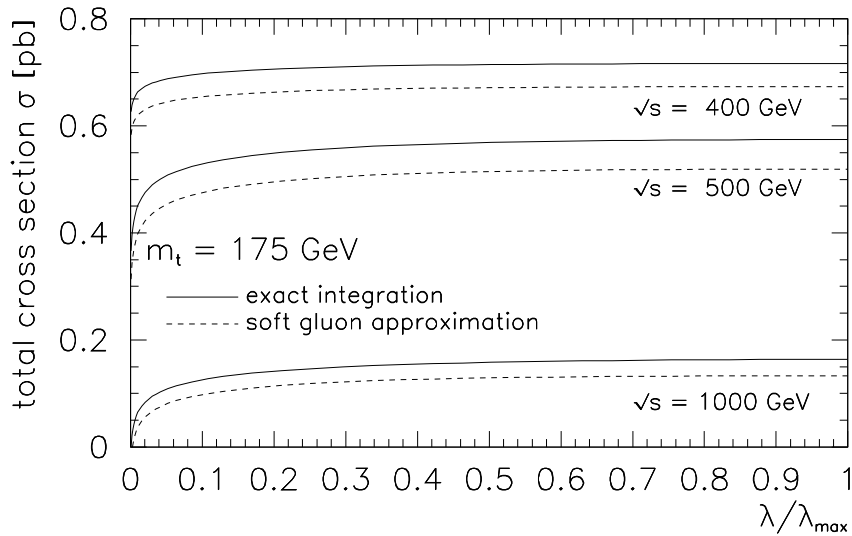}
   \end{center}
   \caption[]{Total cross section for
    $ e^+ e^- \rightarrow t \bar{t} (g) $
    as a function of the scaled gluon energy cut
    $ \lambda / \lambda_{max} $ for different c.m. energies.
    Full line: full calculation; dashed line: soft gluon approximation}
   \label{plot7}
 \end{figure}

 In the soft gluon (or photon) approximation one keeps only the
 soft gluon (or photon) piece in the tree graph contribution but
 includes the full one--loop contribution.
 New structure is thus only generated by the one--loop
 contribution since the soft gluon piece has Born term structure.
 The structure contained in the hard part of the tree graph contribution
 is lost in the soft gluon approximation.
 A brief glance at the corresponding hard photon piece in (\ref{matrix-element})
 shows that it is considerably more difficult to do the analytic phase space
 integration for the hard part than for the soft photon piece.
 In particular, the integration of the hard photon piece has to be done
 separately for each density matrix element.
 Contrary to this, the integration of the soft photon piece
 has to be done only once since the Born term factor can be factored out
 of the integral. Technically, the soft gluon (or photon) approximation
 is much simpler than the full calculation if done analytically. Of course,
 if the calculation is done numerically, the integration of the hard
 gluon (or photon) part causes no additional problems. When the soft gluon
 (or photon) approximation is used this is done at the cost of loosing
 interesting structure contained in the hard gluon (or photon) part.

 In order to be able to judge the quality of the soft gluon 
 approximation we have calculated the rate for
 $ e^+ e^- \rightarrow t \, \bar{t}\, (g) $ and have compared
 the results of the full calculation with the results of the
 soft gluon approximation using different cut--off values for
 the gluon energy \cite{gk03}.
 Fig.~\ref{plot7} shows that the soft gluon result is below
 the full result for the whole range of cut--off values
 $ \lambda = E_g / \sqrt{q^2} $ ($ \lambda_{max} = 1 - 2 m_t/\sqrt{q^2} $).
 For maximal cut--off values $ \lambda = \lambda_{max} $ the
 soft gluon approximation is $ 3.4 \% $, $ 5 \% $ and $ 11.5 \% $ 
 below the full result at $ 400 \GeV $, $ 500 \GeV $ and
 $ 1000 \GeV $, respectively.
 As emphasized above, much of the rate that is being missed
 by the soft gluon approximation has interesting structure.

%%%%%%%%%%%%%%%%%%%%%%%%%%%%%%%%%%%%%%%%%%%%%%%%%%%%%%%%%%%%%%%%%%%%%%%%%%%%%%%%
%
%               Unpolarized Top Decay t->b + W^+
%
%%%%%%%%%%%%%%%%%%%%%%%%%%%%%%%%%%%%%%%%%%%%%%%%%%%%%%%%%%%%%%%%%%%%%%%%%%%%%%%%

 \section{Unpolarized Top Decay \pf $ t \rightarrow b + W^+ $}

 We have already discussed various aspects of the decay
 $ t \rightarrow b + W^+ $ in previous sections.
 In this section we concentrate on unpolarized top decay.
 In particular, we want to discuss the mass dependence
 of the longitudinal piece of the $ W^+ $ boson the measurement
 of which could lead to an independent determination of the mass
 of the top quark.

 The angular decay distribution for unpolarized top decay can be obtained 
 by setting $ P = 0 $ in (\ref{DiffRate}).
 The contribution of the three remaining structure functions
 $ H_L $, $ H_U $ and $ H_F $ can be disentangled by a measurement
 of the shape of the lepton energy spectrum from top decay.
 Given enough data, one can hope to determine the
 longitudinal contribution with $ 1 \% $ accuracy \cite{amidei96}.

\begin{figure}[htbp] %% plot VIII
   \begin{center}
     \includegraphics[width=.60 \textwidth]{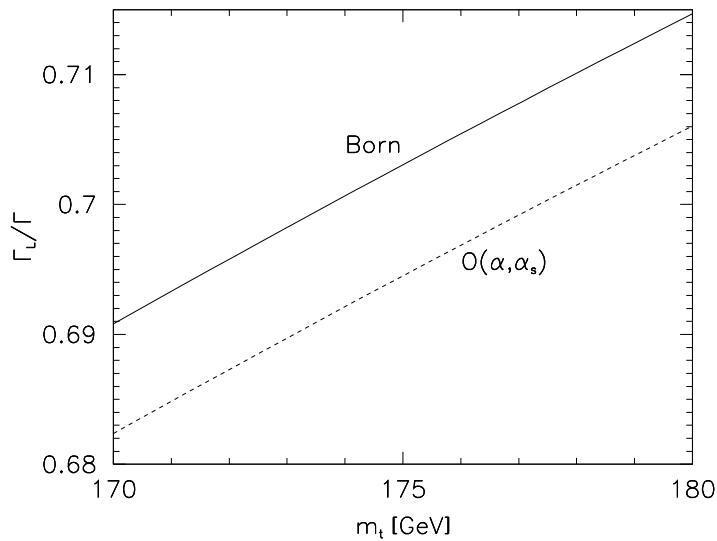}
   \end{center}
   \caption[]{Top mass dependence of the ratio
    $ \Gamma_L / \Gamma $ (full line: LO, dashed line:
    corrections including QCD, electroweak ($ G_F $--scheme),
    finite--width and ($ m_b \ne 0 $) Born term corrections)}
   \label{plot8}
 \end{figure}

 At \LO the mass dependence of the longitudinal contribution is given by 
 $ \Gamma_{L}/\Gamma = 1/(1 + 2(m_W / m_t)^2) $ which gives 
 $ \Gamma_{L}/\Gamma = 0.703 $ using $ m_t = 175 \GeV $ and
 $ m_W = 80.419 \GeV $.
 \NLO corrections to the longitudinal contribution
 were calculated in \cite{fgkm01,fgkm02} (QCD) and in
 \cite{dgkm03} (electroweak and finite width).
 Curiously enough the electroweak and finite width corrections
 tend to cancel each other in the structure functions.      
 In Fig.~\ref{plot8} we show the top mass dependence of
 the ratio $ \Gamma_L / \Gamma $.
 The Born term and the corrected curves are
 practically straight line curves.
 The horizontal displacement of the two
 curves is $ \approx 3.5 \GeV $.
 One would thus make the corresponding mistake in the top mass
 determination from a measurement of $ \Gamma_L / \Gamma $
 if the Born term curve were used instead of the corrected curve.
 If we take $ m_t = 175 \GeV $  as central value,
 a $ 1 \% $ relative error on the measurement of
 $ \Gamma_L / \Gamma $ would allow one to determine
 the top quark mass with $ \approx  3 \GeV $ accuracy.
 Such a top mass measurement would be a welcome alternative to
 the usual invariant mass determination of the top quark mass since
 the $ \Gamma_L / \Gamma $ measurement is a completely independent
 measurement of the top quark mass.
 
%%%%%%%%%%%%%%%%%%%%%%%%%%%%%%%%%%%%%%%%%%%%%%%%%%%%%%%%%%%%%%%%%%%%%%%%%%%%%%%%
%
% The Decay t(up) -> b + H+
%
%%%%%%%%%%%%%%%%%%%%%%%%%%%%%%%%%%%%%%%%%%%%%%%%%%%%%%%%%%%%%%%%%%%%%%%%%%%%%%%%

 \section{The Decay \pf $ t(\uparrow) \rightarrow b + H^+ $}

 Certain extensions of the minimal Standard Model with two
 Higgs--doublets (as e.g. the Minimal Supersymmetric Standard Model)
 contain charged Higgs bosons such that the decay
 $ t(\uparrow) \rightarrow b + H^+ $ would be possible if
 kinematically allowed.
 The Lagrangian for the decay reads

 \begin{equation}
   {\cal L} =
   \bar{\Psi}_{b}(x) 
   (a + b \gamma_5)
   \Psi_{t}(x)
   \phi_{H^+}(x) + \mbox{h.c.}
   \hspace{1mm}. 
 \end{equation}

 \noindent For the differential decay rate one obtains
 
 \begin{equation} % *** differential decay rate *** %
   \label{difrate}
   \frac{d \Gamma}{d \! \cos \theta_P} = \frac{1}{2} \,
   \Big(\Gamma + P \, \Gamma^P \cos \theta_P \Big) = \frac{1}{2} \,
   \Gamma \Big(1 + P \alpha_H \cos \theta_P \Big),
 \end{equation}

 \noindent where $ \theta_P $ is the angle between the
 polarization vector of the top quark and the Higgs.
 The $ \cos \theta_P $--dependence in (\ref{difrate})
 is determined by the asymmetry parameter
 $ \alpha_H = \Gamma / \Gamma_P $.
 At \LO one obtains $ ( m_b=0) $
  
 \begin{equation} % *** asymmetry parameter *** %
   \label{asym} \alpha_H = \frac{2 a b}{a^2 + b^2}.
 \end{equation}

 At this point we want to emphasize that one needs the polarization
 information contained in (\ref{difrate}) to be able to determine the
 relative sign of the two coupling constants $ a $ and $ b $.
 The rate is proportional to $ (a^2 + b^2) $ and is therefore
 insensitive to the relative sign of the coupling constants.
 
 We now specify to the so called model~2 where the
 coupling structure is expressed in terms of
 $ \tan \beta = v_2/v_1 $, and where $ v_1 $ and $ v_2 $
 are the vacuum expectation values of the two
 neutral components of the two Higgs doublets.
 In model~2 the coupling constants are given by 
 
 \begin{eqnarray} % *** model 2 *** %
  a & = & \frac{g_w}{2 \sqrt{2} m_W}
  V_{tb}(m_t \cot \beta + m_b \tan \beta), \\[3mm] 
  b & = & \frac{g_w}{2 \sqrt{2} m_W}
  V_{tb}(m_t \cot \beta - m_b \tan \beta).
 \end{eqnarray}

 \begin{figure}[htbp] %% plot IX
   \begin{center}
     \includegraphics[width=.60 \textwidth]{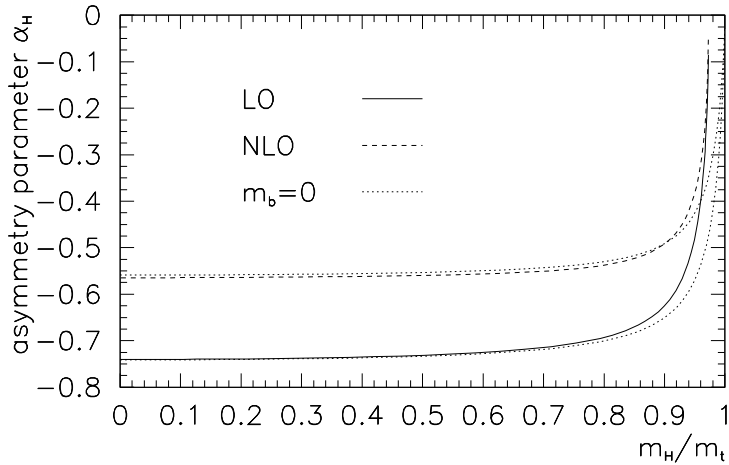}
   \end{center}
   \caption[]{Asymmetry parameter $ \alpha_H $ for model~2 with
    $ m_b = 4.8 \GeV $ and $ m_t = 175 \GeV $ (LO: full line,
    NLO: dashed line) as function of $ m_{H^+}/m_t $ and
    $ \tan \beta = 10 $.
    The barely visible dotted lines show the corresponding
    $ m_b \rightarrow 0 $ curves.}
   \label{plot9}
 \end{figure}

 \noindent The weak coupling factor $ g_w $ is related to
 the usual Fermi coupling constant $ G_F $ by
 $ g_w = 2 m _W \sqrt[4]{2} \sqrt{G_F}$.
 In Fig.~\ref{plot9} we show a plot of the asymmetry parameter
 $ \alpha_H $ as a function of the mass ratio $ m_H^+/m_t $ for
 a fixed value of $ \tan \beta = 10 $.
 The asymmetry parameter is large and negative
 over most of the range of mass ratios.
 The radiative corrections can be seen to be quite substantial.  

%%%%%%%%%%%%%%%%%%%%%%%%%%%%%%%%%%%%%%%%%%%%%%%%%%%%%%%%%%%%%%%%%%%%%%%%%%%%%%%%
%
% Goldstone Equivalence Theorem 
%
%%%%%%%%%%%%%%%%%%%%%%%%%%%%%%%%%%%%%%%%%%%%%%%%%%%%%%%%%%%%%%%%%%%%%%%%%%%%%%%%

 \section{Goldstone Equivalence Theorem}

 Consider the decay $ t \rightarrow b + J_{\mu}^+(q^2) $ where the current 
 $ J_{\mu}^{+}(q^2) $ can be thought of as representing an off--shell $ W^+ $
 containing spin $ 1 $ and spin $ 0 $ pieces.
 Taking $ q^{\mu} = (q_0; 0, 0, |\vec{q}|) $ the longitudinal polarization
 vector of the spin $ 1 $ piece is given by
 $ \epsilon^{\mu}(L)= (|\vec{q}|; 0, 0, q_0) / \sqrt{q^2} $
 while the scalar polarization vector reads
 $ \epsilon^{\mu}(S) = q^{\mu} / \sqrt{q^2} $.
 In the limit $ \sqrt{q^2} / m_t \rightarrow 0 $
 the longitudinal polarization vector becomes
 increasingly parallel to its momentum.
 In fact, one finds
 $ \epsilon^{\mu}(L) = q^{\mu} / \sqrt{q^2} + O(\sqrt{q^2} / q_0) $.
 In this limit the amplitude $ A(L) $ dominates over the transverse
 amplitudes $ A(T) $ since $ A(T) / A(L) \sim \sqrt{q^2}/q_0 $.
 At the same time one finds $ A(S) \sim A(L) $ in this limit.
 This provides a useful check on the high energy limits
 of the relevant unpolarized and polarized rate functions.

 Next consider the scalar projection of the
 $ t \rightarrow b $ current transition at lowest order.
 Using the Dirac equation one obtains ($ q = p_t - p_b $)

 \begin{eqnarray}
   \label{scalarproj}
   q_{\mu} \bar{u}_b \gamma^{\mu}
   (\1 - \gamma_5) u_t & = &
   (p_t - p_b)_{\mu} \bar{u}_b
   \gamma^\mu (\1 - \gamma_5)
   u_t \nonumber \\[2mm] & = &
   \bar{u}_b (m_t (\1 + \gamma_5) +
   m_b u_t(\1 - \gamma_5)) u_t.
 \end{eqnarray}  

 \noindent The second line of (\ref{scalarproj}) has the coupling structure
 of a Goldstone boson in the Standard Model.
 It is a very instructive exercise to follow this argument
 through to one--loop order including renormalization \cite{Czarnecki:yz}.
 In fact, this generalizes to any order in perturbation theory:
 the scalar piece of the charged transition current has the
 coupling structure of a Goldstone boson.
 Together with the above statement that the longitudinal rate
 dominates at high energies and that it is equal to the scalar rate
 in this very limit one arrives at the Goldstone Boson Equivalence Theorem.
 In the high energy limit the coupling of a gauge vector boson to a
 fermion current is equivalent to that of a Goldstone boson.
 This simplifying feature has frequently been used in the literature
 to dramatically reduce the effort needed in the computation of
 high energy processes involving vector gauge bosons.
 For example, the Goldstone equivalence theorem was used in
 \cite{Korner:1995xd} to calculate the bosonic two--loop
 electroweak radiative corrections to the decay
 $ H \rightarrow \gamma + \gamma $ in the limit of a large Higgs mass.
 The same approximation was used to calculate the dominant
 bosonic two--loop electroweak radiative corrections to the
 decay of a heavy Higgs into pairs of $ W $ and $ Z $
 gauge bosons \cite{Frink:1996sv}.

 As concerns the coupling of charged Higgs bosons in
 $ t \rightarrow b + H^+ $ discussed in Sect.~10 the
 coupling structure of the charged Higgs is equal to that
 of a Goldstone boson in the Standard Model if one specifies
 to the so called model $ 1 $ with
 $ \cot \beta = 1 $ (see e.g. \cite{km03}).
 In this case one has

 \begin{equation}  
   \Gamma_{t \rightarrow b + H^+} \sim
   \Gamma_{t \rightarrow b + W^+} \hspace{1cm}  \mathrm{as}
   \hspace{1cm} m_{W^+}, m_{H^+} / m_t \rightarrow 0.
 \end{equation} 

 The same relation holds true for the corresponding polarized top rates.
 All of the statements made in this section have been explicitly verified
 using our \NLO results on $ t \rightarrow b + W^+ $ \cite{fgkm02} and
 $ t \rightarrow b + H^+ $ \cite{km03}.

%%%%%%%%%%%%%%%%%%%%%%%%%%%%%%%%%%%%%%%%%%%%%%%%%%%%%%%%%%%%%%%%%%%%%%%%%%%%%%%%
%
% Leptonic Decays of the mu and the tau and
% anomalous helicity flip contributions
%
%%%%%%%%%%%%%%%%%%%%%%%%%%%%%%%%%%%%%%%%%%%%%%%%%%%%%%%%%%%%%%%%%%%%%%%%%%%%%%%%

 \section{Leptonic Decays of the \pf $ \mu $ and the $ \tau $ and
 Anomalous Helicity Flip Contributions}
 
 We have already discussed various aspects of the leptonic decays of the
 $ \mu $ and the $ \tau $ lepton in previous sections.
 A complete solution to the problem has been given in \cite{fgkm03electron}
 where, for the first time, all mass and polarization effects have been
 included at {\sl NLO}.
 Partial \NLO results on polarization can be found in \cite{fs74,arbuzov02}.
 In this section we concentrate on one aspect of the problem, namely on the
 so--called anomalous contribution that, in the chiral limit, flips the
 helicity of the final--state lepton at {\sl NLO}.

 Collinear photon emission from a massless fermion line can flip the
 helicity of the massless fermion contrary to naive expectation.
 This has been discussed in a variety of physical contexts.
 This is a ``$ m_e/m_e $'' effect where the $ m_e $ in the numerator
 is a spin flip factor and the $ m_e $ in the denominator arises from
 the collinear configuration.
 In the limit $ m_e \rightarrow 0 $ the helicity flip contribution
 survives whereas it is not seen in massless QED. 
 
 We shall discuss this phenomenon in the context of the left--chiral
 $ \mu \rightarrow e $ transition.
 At the Born term level an electron emerging from a
 weak $ (V - A) $ vertex is purely left--handed in the limit $ m_e = 0 $.
 Naively, one would expect this to be true also at $ O(\alpha) $ or
 at any order in $ \alpha $ because in massless QED photon
 emission from the electron is helicity conserving.
 Let us make this statement more precise by looking at the string
 of $ \gamma $--matrices between the initial state $ \mu $ spinor
 and final state $ e $ antispinor of the left--chiral
 $ \mu \rightarrow e $ transition.
 Using $ (1 - \gamma_5) = (1 - \gamma_5)(1 - \gamma_5)/2 $,
 and the fact that in massless QED every photon emission brings
 in two $ \gamma $--factors (one from the vertex and one from the
 fermion propagator) one finds

 \begin{equation}
   \bar{u}_e \Gamma^{(n)}
   \gamma^\mu (\1 - \gamma_5) u_\mu =
   \frac{1}{2} \bar{u}_e
   (\1 + \gamma_5) \Gamma^{(n)}
   \gamma^{\mu} (\1 - \gamma_5) u_\mu,
 \end{equation}

 \noindent where $ \Gamma^{(n)} $ stands for the (even) $ \gamma $
 matrix string brought in by the emission of $ n $ photons.
 We have commuted the left--chiral factor $ (1 - \gamma_5) $
 to the left end where it then projects out the helicity state
 $ \lambda_e = -1/2 $ from the electron antispinor thus proving
 the above assertion.

 Let us take a closer look at the anomalous helicity flip contribution
 in leptonic $ \mu \rightarrow e $ decays by considering the
 unnormalized density matrix element $ \rho_{++} $ of the
 final state electron which is obtained by setting
 $ \cos \theta = 1 $ in (\ref{diffrate})
 (remember that $ G_5 $ vanishes for $ m_e \rightarrow 0 $
 and $ G_6 = 0 $ in the Standard Model).
 One has 

 \begin{equation} % *** no--flip and flip  *** %
   \label{plusplus1}
   \frac{d \Gamma^{(+)}}{dx \, d\!\cos \theta_P } =
   \frac{1}{2} \beta x \, \Gamma_0
   \Big( ( G_1 + G_3) + ( G_2 + G_4) P \cos \theta_P \Big). 
 \end{equation}

 Contrary to naive expectations one finds non--vanishing
 right--handed $ (+) $ contributions which survive the
 $ m_e \rightarrow 0 $ limit when one takes the
 $ m_e \rightarrow 0 $ limit of the \NLO contributions to
 (\ref{plusplus1}) \cite{fgkm03electron}.
 In fact, one finds 

 \begin{equation}
   \label{plusplus2}
   \frac{d \Gamma^{(+)}}{dx \, d\!\cos \theta_P } =
   \frac{\alpha}{12 \pi} \Gamma_0
   \Big( \Big[ (1 - x)^2 (5 - 2 x) \Big] -
   \Big[ (1 - x)^2 (1 + 2 x) \Big] P \cos \theta_P \Big).
 \end{equation}

 The result is rather simple.
 In particular, it does not contain any logarithms or dilogarithms.
 The simplicity of the right--handed contribution becomes manifest
 in the equivalent particle description of $ \mu $--decay where,
 in the peaking approximation, $ \mu $--decay is described by the
 two--stage process $ \mu^- \rightarrow e^- $ followed by the
 branching process $ e^- \rightarrow e^- + \gamma $ characterized
 by universal splitting functions $ D_{n\!f / h\!f}(z) $ \cite{falk94}.
 The symbols $ n\!f $ and $ h\!f $ stand for a helicity non--flip and
 helicity flip of the helicity of the electron.
 In the splitting process $ z $ is the
 fractional energy of the emitted photon.
 The off--shell electron in the propagator is replaced by an
 equivalent on--shell electron in the intermediate state.
 Since the helicity flip contribution arises entirely
 from the collinear configuration it can be calculated
 in its entirety using the equivalent particle description.

 The helicity flip splitting function is given by
 $ D_{h\!f}(z) = \alpha z/(2\pi) $, where
 $ z = k_0 / E' = (E' - E)/ E' = 1 - x/x' $,
 and where $ k_0 $ is the energy of the emitted photon.
 $ E' $ and $ E $ denote the energies of the
 initial and final electron in the splitting process.
 The helicity flip splitting function has to be folded with the
 appropriate $ m_e = 0 $ Born term contribution.
 The lower limit of the folding integration is determined
 by the soft photon point where $ E' = E $.
 The upper limit is determined by the maximal energy of the
 initial electron $ E' = m_{\mu}/{2} $. 
 One obtains 

 \begin{eqnarray} %% anomalous contribution
   \label{plusplus3}
   \frac{d \Gamma^{(+)}}{dx \, d\!\cos \theta_P } & = &
   \frac{\alpha}{2 \pi}
   \int_x^1 dx' \frac{1}{x'}
   \frac{d \Gamma^{{\mathrm {Born}}; (-)} (x')}
    {dx' \, d\!\cos \theta_P }
    (1 - \frac{x}{x'}) \nonumber \\[2mm] & = &
   \frac{\alpha}{2 \pi} \Gamma_0
   \int_x^1 dx' (x' - x)
   \Big( (3 - 2 x') + (1 - 2 x')
    P \cos \theta_P \Big) \nonumber \\[2mm] & = &
   \frac{\alpha}{12 \pi} \Gamma_0
   \Big( (1 - x)^2 (5 - 2 x) - (1 - x)^2 (1 + 2 x)
    P \cos \theta_P \Big),
 \end{eqnarray}

 \noindent which exactly reproduces
 the result (\ref{plusplus2}).
   
 Numerically, the flip spectrum function is rather small
 compared to the $ O(\alpha_s) $ no--flip spectrum function.
 However, when averaging over the spectrum the ratio of the
 $ O(\alpha_s) $ flip and no--flip contributions amounts to a
 non--negligible $ (- 12 \%) $, due to cancellation effects in the
 $ O(\alpha_s) $ no--flip contribution.

%%%%%%%%%%%%%%%%%%%%%%%%%%%%%%%%%%%%%%%%%%%%%%%%%%%%%%%%%%%%%%%%%%%%%%%%%%%%%%%%
%
% Summary; Concluding Remarks 
%
%%%%%%%%%%%%%%%%%%%%%%%%%%%%%%%%%%%%%%%%%%%%%%%%%%%%%%%%%%%%%%%%%%%%%%%%%%%%%%%%

 \section{Summary and Concluding Remarks}

 We have discussed \NLO corrections to a multitude of polarization observables
 in different processes including nonzero mass effects.
 The results are available in compact analytical form.
 They are ready for use in physics simulation programs which
 are reliable even at corners of phase space where mass effects
 become important.
 Present and planned experiments (TEVATRON Run2, LHC, BELLE, BABAR,
 $ \tau $--charm factories) will be sensitive to these
 \NLO Standard Model effects.
 Standard Model \NLO corrections to polarization observables
 are needed as background for possible new physics contributions.
 Last but not least the \NLO results are needed for
 \NLO sum rule analysis' involving polarization observables.
 All calculations are based on the same one--loop and
 tree--graph matrix elements.
 Results on rates agree with previous calculations.
 Results on polarization observables agree with
 previous calculations where available.
 We have checked various limits and found agreement with
 previous mass zero calculations, the Goldstone boson
 equivalence theorem and the equivalent particle description of the
 anomalous helicity flip contribution.
 Because of the various checks and the fact that we have
 essentially used one set of matrix elements as input to the
 calculations we feel quite confident that our results are correct.

%%%%%%%%%%%%%%%%%%%%%%%%%%%%%%%%%%%%%%%%%%%%%%%%%%%%%%%%%%%%%%%%%%%%%%%%%%%%%%%%
%
% acknowledgements
%
%%%%%%%%%%%%%%%%%%%%%%%%%%%%%%%%%%%%%%%%%%%%%%%%%%%%%%%%%%%%%%%%%%%%%%%%%%%%%%%%

 \vspace{10mm} {\noindent \bf Acknowledgements:}
 We would like to thank the organizers of this meeting, D.~Blaschke,
 M.A.~Ivanov and S.~Nedelko for providing a most graceful conference
 setting. 
 We acknowledge informative discussions and e--mail
 exchanges with A.B.~Arbuzov, F.~Berends, W.~van Neerven,
 F.~Scheck, K.~Schilcher, H.~Spiesberger and O.V.~Teryaev.
 We would like to thank B.~Lampe, M.~Fischer, S.~Groote
 and H.S.~Do for their collaborative effort.
 M.C.~Mauser is supported by the DFG (Germany)
 through the Graduiertenkolleg ``Eichtheorien''
 at the University of Mainz.

%%%%%%%%%%%%%%%%%%%%%%%%%%%%%%%%%%%%%%%%%%%%%%%%%%%%%%%%%%%%%%%%%%%%%%%%%%%%%%%%
%
% bibliography
%
%%%%%%%%%%%%%%%%%%%%%%%%%%%%%%%%%%%%%%%%%%%%%%%%%%%%%%%%%%%%%%%%%%%%%%%%%%%%%%%%

\end{document}